\documentclass[journal]{IEEEtran}
\ifCLASSOPTIONcompsoc
\usepackage[caption=false,font=normalsize,labelfon
t=sf,textfont=sf]{subfig}
\else
\usepackage[caption=false,font=footnotesize]{subfi
g}
\fi

\usepackage{amsmath,amssymb}
\usepackage{multicol}
\usepackage{graphicx,graphics,color,psfrag}
\usepackage{cite,balance}
\usepackage{algorithm}
\usepackage{accents}

\usepackage{bm}
\usepackage{url}
\usepackage{algorithmic}
\usepackage[english]{babel}
\usepackage{multirow}
\usepackage{enumerate}
\usepackage{cases}
\usepackage{stfloats}
\usepackage{dsfont}
\usepackage{color,soul}
\usepackage{amsfonts}
\usepackage{tcolorbox}
\usepackage{amsmath}
\usepackage{float}

\usepackage{cite,graphicx,amsmath,amssymb}
\usepackage{fancyhdr}
\usepackage{hhline}
\usepackage{graphicx,graphics}
\usepackage{array,color}
\usepackage{amsmath}
\usepackage{stfloats}
\usepackage[flushleft]{threeparttable}

\newtheorem{proposition}{Proposition}

\newtheorem{lemma}{Lemma}

\ifCLASSINFOpdf

\else

\fi

\begin{document}

\title{Intelligent Reflecting Surface Enabled Sensing: Cram\'er-Rao Bound Optimization}

\author{Xianxin~Song,~\IEEEmembership{Student Member,~IEEE}, Jie~Xu,~\IEEEmembership{Senior Member,~IEEE}, Fan~Liu,~\IEEEmembership{Member,~IEEE}, \\Tony~Xiao~Han,~\IEEEmembership{Senior Member,~IEEE}, and  Yonina C.~Eldar,~\IEEEmembership{Fellow,~IEEE}
\thanks{Part of this paper has been submitted to IEEE Globecom 2022 \cite{xianxin}.
}
\thanks{Xianxin Song and Jie Xu are with the School of Science and Engineering (SSE), Future Network of Intelligence Institute (FNii),
and Guangdong Provincial Key Laboratory of Future Networks of Intelligence, The Chinese University of Hong Kong (Shenzhen), Shenzhen 518172, China (e-mail: xianxinsong@link.cuhk.edu.cn, xujie@cuhk.edu.cn). Jie Xu is the corresponding author.}
\thanks{Fan Liu is with the Department of Electrical and Electronic Engineering, Southern University of Science and Technology, Shenzhen 518055, China (email: liuf6@sustech.edu.cn).}
\thanks{Tony Xiao Han is with the Wireless Technology Lab, 2012 Laboratories, Huawei, Shenzhen 518129, China (e-mail: tony.hanxiao@huawei.com).}
\thanks{Yonina C. Eldar is with the Faculty of Mathematics and Computer Science, Weizmann Institute of Science, Rehovot 7610001, Israel (e-mail: yonina.eldar@weizmann.ac.il).}}

\maketitle

\begin{abstract}
This paper investigates intelligent reflecting surface (IRS) enabled non-line-of-sight (NLoS) wireless sensing, in which an IRS is dedicatedly deployed to assist an access point (AP) to sense a target at its NLoS region. It is assumed that the AP is equipped with multiple antennas and the IRS is equipped with a uniform linear array. We consider two types of target models, namely the point and extended targets, for which the AP aims to estimate the target's direction-of-arrival (DoA) and the target response matrix with respect to the IRS, respectively, based on the echo signals from the AP-IRS-target-IRS-AP link. Under this setup, we jointly design the transmit beamforming at the AP and the reflective beamforming at the IRS to minimize the Cram\'er-Rao bound (CRB) on the estimation error. Towards this end, we first obtain the CRB expressions for the two target models in closed form. It is shown that in the point target case, the CRB for estimating the DoA depends on both the transmit and reflective beamformers; while in the extended target case, the CRB for estimating the target response matrix only depends on the transmit beamformers. Next, for the point target case, we optimize the joint beamforming design to minimize the CRB, via alternating optimization, semi-definite relaxation, and successive convex approximation. For the extended target case, we obtain the optimal transmit beamforming solution to minimize the CRB in closed form. Finally, numerical results show that for both cases, the proposed designs based on CRB minimization achieve improved sensing performance in terms of mean squared error, as compared to other traditional schemes.
\end{abstract}


\IEEEpeerreviewmaketitle

\section{Introduction}
Integrating wireless (radar) sensing into future beyond-fifth-generation (B5G) and six-generation (6G) wireless networks as a new functionality has attracted growing research interest to support various environment-aware applications such as industrial automation, auto-driving, and remote healthcare (see, e.g., \cite{liu2020joint,liu2021integrated,9540344} and the references therein). Conventionally, wireless sensing relies on line-of-sight (LoS) links between the access points (APs) and the sensing targets, such that the sensing information of targets can be extracted based on the target echo signals\cite{liu2021integrated}. However, in practical scenarios with dense obstructions, sensing targets are likely to be located at the non-LoS (NLoS) region of APs, where conventional LoS sensing may not be applicable in general. Therefore, how to realize NLoS sensing in such scenarios is a challenging task.

Motivated by its success in wireless communications \cite{8811733,9326394,9122596}, intelligent reflecting surface (IRS) or reconfigurable intelligent surface (RIS) has become a viable new solution to overcome this issue (see, e.g., \cite{aubry2021reconfigurable,shao2022target,9361184,9454375,Stefano,song2021joint,9364358,9769997}). By properly deploying IRSs around the AP to reconfigure the radio propagation environment, virtual LoS links are established between the AP and the targets in its NLoS region, such that the AP can perform NLoS target sensing based on the echo signals from  AP-IRS-target-IRS-AP links. To combat the severe signal propagation loss over such triple-reflected links, the IRS can adaptively control the phase shifts at reflecting elements, such that the reflected signals are beamed towards desired target directions to enhance sensing performance.

There have been several prior works investigating IRS-enabled wireless sensing \cite{aubry2021reconfigurable,shao2022target,9361184,9454375,Stefano} and IRS-enabled integrated sensing and communications (ISAC) \cite{song2021joint,9364358,9769997}, respectively. The work \cite{aubry2021reconfigurable} presented the NLoS radar equation based on the AP-IRS-target-IRS-AP link and evaluated the resultant sensing performance in terms of signal-to-noise ratio (SNR) and signal-to-clutter ratio (SCR). In \cite{shao2022target}, the authors considered an IRS-enabled bi-static target estimation, where dedicated sensors were installed at the IRS for estimating the direction of its nearby target through the AP-IRS-target-sensors link. Under this setup, the authors optimized the IRS's reflective beamforming to maximize the average received signal power at the sensors and applied the multiple signal classification (MUSIC) algorithm for estimating targets.
In \cite{9361184,9454375,Stefano}, the authors considered IRS-enabled target detection, in which the IRS's passive beamforming was optimized to maximize the target detection probability subject to a fixed false alarm probability constraint. In \cite{song2021joint}, the authors considered an IRS-enabled ISAC system with one base station (BS), one communication user (CU), and multiple targets, in which the IRS's minimum beampattern gain towards the desired sensing angles was maximized by jointly optimizing the transmit beamforming at the BS and the reflective beamforming at the IRS, subject to the minimum SNR requirement at the CU. In \cite{9364358,9769997}, the authors considered an IRS-enabled ISAC system, in which the SNR of radar was maximized by joint beamforming design while ensuring the quality-of-service requirement at the CU.

In those prior works on IRS-enabled sensing and IRS-enabled ISAC, the sensing SNR (or beampattern gain) and the target detection probability have been widely adopted as the sensing performance measure for joint beamforming optimization. By contrast, Cram\'er-Rao bound (CRB) is another important sensing performance measure, especially for target estimation tasks, which provides a lower bound on the variance of unbiased parameter estimators. In prior studies on wireless sensing\cite{kay1993fundamentals,bekkerman2006target,4359542} and ISAC\cite{9652071,rate_CRB} without IRS, CRB has been widely adopted as the design objective for sensing performance optimization. Nevertheless, to our best knowledge, how to analyze the CRB performance for NLoS target estimation through the AP-IRS-target-IRS-AP link and optimize such performance by joint transmit and reflective beamforming design is still an uncharted area.

This paper considers an IRS-enabled NLoS wireless sensing system, which consists of one AP with multiple antennas, one IRS with a uniform linear array (ULA), and one target at the NLoS region of the AP. In particular, we focus on the narrowband transmission, in which a general multi-path channel model (including both LoS and NLoS paths) is considered for the AP-IRS link. It is assumed that the AP perfectly knows the channel state information (CSI) of the AP-IRS link, and the AP needs to estimate the target parameters based on the echo signals from the AP-IRS-target-IRS-AP link. We consider two types of target models \cite{259642,9652071,4200705,8579200}, namely the point and extended targets, respectively.
For the point target case, the AP needs to estimate the target's direction-of-arrival (DoA) with respect to (w.r.t.) the IRS. For the extended target case, the AP needs to estimate the complete target response matrix w.r.t. the IRS (or equivalently the cascaded IRS-target-IRS channel matrix). We aim to minimize the CRB for parameters estimation, by jointly optimizing the transmit beamforming at the AP and the reflective beamforming at the IRS.  The main results are listed as follows.
\begin{itemize}
\item First, we obtain the closed-form CRB expressions for estimating the DoA and the target response matrix in the point and extended target cases, respectively. For the point target case, it is shown that the CRB depends on both the transmit and reflective beamformers, and the target's DoA is only estimable when the rank of the AP-IRS channel matrix is higher than one or equivalently there are at least two signal paths in that channel. By contrast, for the extended target case, it is revealed that the target response matrix is only estimable when the rank of the AP-IRS link is higher than the number of reflecting elements at the IRS, and the resultant CRB only depends on the transmit beamformers at the AP.
\item Next, we minimize the obtained CRB by jointly optimizing the transmit beamforming at the AP and the reflective beamforming at the IRS, subject to a maximum power constraint at the AP. For the point target case, the formulated CRB minimization problem is non-convex and difficult to solve. To resolve this issue, we present an efficient algorithm via alternating optimization, semi-definite relaxation (SDR), and successive convex approximation (SCA). For the extended target case, the resultant CRB minimization problem is convex, in which only the transmit beamforming vectors at the AP are optimization variables. We obtain the closed-form optimal transmit beamforming solution, in which the singular value decomposition (SVD) is first implemented to diagonalize the AP-IRS channel into parallel subchannels, and then the channel amplitude inversion power allocation is performed over these subchannels.
\item Finally, we present numerical results to validate the performance of our proposed designs. It is shown that the for the point target case, joint beamforming design based on CRB minimization achieves improved sensing performance in terms of mean squared error (MSE), as compared to the traditional schemes with SNR maximization, separate beamforming designs, and isotropic transmission. For the extended target case, the proposed transmit beamforming design is shown to outperform the traditional isotropic transmission scheme. We also compare the CRB performance versus the  MSE achieved by maximum likelihood estimation (MLE). For the point target case, the MSE is shown to converge towards the CRB when the SNR becomes sufficiently high. For the extended target case, the MSE equals to the CRB.
\end{itemize}

The remainder of this paper is organized as follows. Section II introduces the system model of our considered IRS-enabled NLoS wireless sensing system. Section III presents the closed-form CRB expressions for estimating the DoA and the target response matrix in the point and extended target cases, respectively. Section IV minimizes the CRB for estimating the DoA in the point target case by jointly optimizing the transmit beamforming at the AP and the reflective beamforming at the IRS. Section V minimizes the CRB for estimating the target response matrix in the extended target case by optimizing the transmit beamforming at the AP. Finally, Section VI provides numerical  results, followed by the conclusion in Section VII.

\textit{Notations:}
Boldface letters refer to vectors (lower case) or matrices (upper case). For a square matrix $\mathbf S$, $\mathrm {tr}(\mathbf S)$ and $\mathbf S^{-1}$ denote its trace and inverse, respectively, and $\mathbf S \succeq \mathbf{0}$ means that $\mathbf S$ is positive semi-definite. For an arbitrary-size matrix $\mathbf M$, $\mathrm {rank}(\mathbf M)$, $\mathbf M^*$, $\mathbf M^{\mathrm {T}}$, and $\mathbf M^{\mathrm {H}}$ are its rank, conjugate, transpose, and conjugate transpose, respectively. The matrix $\mathbf I_m$ is an identity matrix of dimension $m$. We use $\mathcal{C N}(\mathbf{x}, \mathbf{\Sigma})$ to denote the distribution of a circularly symmetric complex Gaussian (CSCG) random vector with mean vector $\mathbf x$ and covariance matrix $\mathbf \Sigma$, and $\sim$ to denote ``distributed as''. The spaces of $x \times y$ complex and real matrices are denoted by $\mathbb{C}^{x \times y}$ and  $\mathbb{R}^{x \times y}$, respectively. The real and imaginary parts of a complex number are denoted by $\mathrm{Re}\{\cdot\}$ and $\mathrm{Im}\{\cdot\}$, respectively.
The symbol $\mathbb{E}(\cdot)$ is the statistical expectation, $\|\cdot\|$ stands for the Euclidean norm, $|\cdot|$ for the magnitude of a complex number, $\mathrm {diag}(a_1,...,a_N)$ for a diagonal matrix with diagonal elements $a_1,...,a_N$, $\otimes$ for the Kronecker product,  $\mathrm{vec}(\cdot)$ for the vectorization operator, and $\mathrm {arg}(\mathbf x)$ for a vector with each element being the phase of the corresponding element in $\mathbf x$.

\section{System Model}

\begin{figure}[t]
\centering
\subfloat[Point target case]{\includegraphics[width=2.5in]{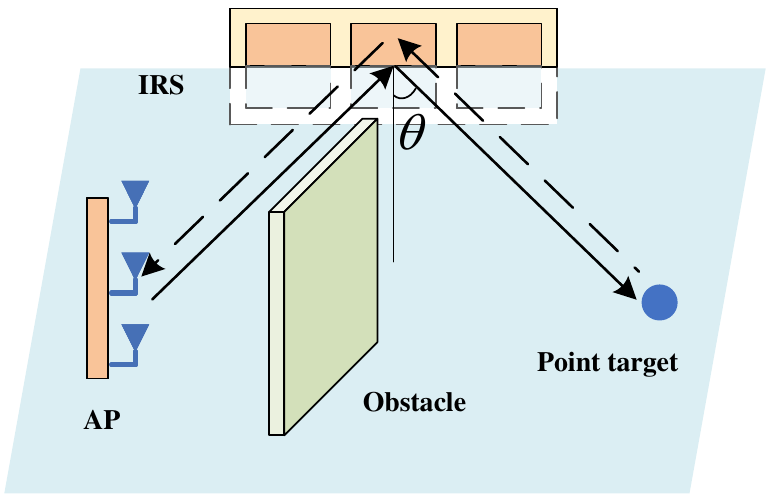}%
\label{Point target}}\\
\subfloat[Extended target case]{\includegraphics[width=2.5in]{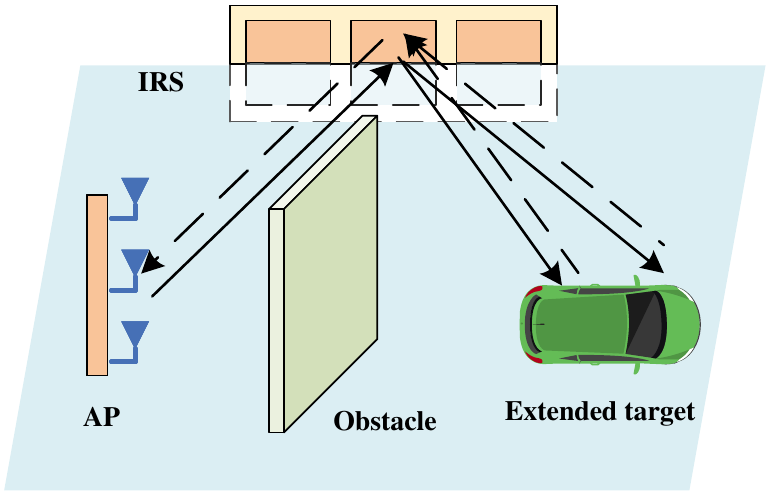}%
\label{Extended target}}
\caption{System model of IRS-enabled sensing.}
\label{system_model}
\end{figure}

We consider an IRS-enabled NLoS wireless sensing system as shown in Fig. \ref{system_model}, which consists of one AP with $M>1$ antennas, one ULA-IRS with $N>1$ reflecting elements, and one target at the NLoS region of the AP (i.e., the LoS link between the AP and the target is blocked by obstacles such as buildings and trees). The IRS is deployed to create a virtual LoS link to facilitate the target sensing. In this case, the AP transmits sensing signals and then estimates the target's parameters w.r.t. the IRS based on the echo signals from the AP-IRS-target-IRS-AP link. The target estimation should be implemented at the AP instead of the IRS, as the IRS is normally a passive device without the capability of signal processing.

First, we consider the transmit and reflective beamforming at the AP and the IRS, respectively. Let $\mathbf x(t)$ denote the transmitted signal by the AP at time slot $t$ and $T$ the radar dwell time. The sample coherence matrix of the transmitted signal is
\begin{equation}
\mathbf R_x=\frac{1}{T}\sum_{t=1}^T \mathbf x(t)\mathbf x(t)^\mathrm{H},
\end{equation}
which corresponds to the transmit beamforming vectors to be optimized. Suppose that $\mathrm{rank}(\mathbf R_x) = k$ and the eigenvalue decomposition (EVD) of $\mathbf R_x$ is given by 
\begin{equation}\label{eq:coherence_matrix_to_vector}
\mathbf R_x=\mathbf W \mathbf \Lambda \mathbf W^\mathrm{H}, 
\end{equation}
where $\mathbf \Lambda = \mathrm{diag}(\lambda_1, ..., \lambda_M)$ and $\mathbf W = [\mathbf w_1,... ,\mathbf w_M], $ with $\lambda_1 \ge ... \ge \lambda_k > \lambda_{k+1}= ... =\lambda_M= 0$ and $\mathbf W  \mathbf W^\mathrm{H} = \mathbf W^\mathrm{H} \mathbf W= \mathbf{I}_M$. This means that there are a number of $k$ sensing beams transmitted by the AP, each of which is denoted by $\sqrt{\lambda_i}\mathbf w_i, i\in\{1, ..., k\}$ (see, e.g., \cite{hua}). As for the reflective beamforming, we consider that the IRS can only adjust the phase shifts of its reflecting elements \cite{8811733}. Let $\mathbf v = [e^{j\phi_1},\ldots,e^{j\phi_{N}}]^\mathrm{T}$ denote the reflective beamforming vector at the IRS, with $\phi_n \in (0, 2\pi]$ being the phase shift of element $n \in \{1,...,N\}$ at the IRS, which is optimized later to enhance the sensing performance.

Next, we introduce the channel models. We consider a general multi-path channel model for the AP-IRS link. Accordingly, let $\mathbf{G} \in \mathbb{C}^{N \times M}$ denote the associated channel matrix of the AP-IRS link, where $\mathrm{rank}(\mathbf{G}) \ge 1$ holds in general. Let $\mathbf H$ denote the target response matrix w.r.t. the IRS (or equivalently the cascaded IRS-target-IRS channel), which can be expressed in the following by considering two specific target models.
\subsubsection{Point Target Model}
When the spatial extent of the target is small, the incident signal is reflected by the target from a singular scatterer \cite{259642,9652071}. The target response matrix w.r.t. the IRS is modeled as
\begin{equation}\label{eq:channel_point}
\mathbf H = \alpha\mathbf a(\theta)\mathbf a^\mathrm{T} (\theta),
\end{equation}
where $\alpha \in \mathbb{C}$ denotes the complex-valued channel coefficient dependent on the target's radar cross section (RCS) and the round-trip path loss of the IRS-target-IRS link, and $\mathbf a(\theta)$ denotes the steering vector at the IRS with angle $\theta$, i.e.,
\begin{equation}\label{eq:steering_vector}
\mathbf a(\theta) = [1,e^{j2\pi\frac{d_\text{IRS}\sin \theta}{\lambda_\text{R}}},...,e^{j2\pi\frac{ (N-1)d_\text{IRS}\sin \theta}{\lambda_\text{R}}}]^\mathrm{T},
\end{equation}
with $\theta$ denotes the target's DoA w.r.t. the IRS. In \eqref{eq:steering_vector},  $d_\text{IRS}$ denotes the spacing between consecutive reflecting elements at the IRS, and $\lambda_\text{R}$ denotes the carrier wavelength. In this case, the target's DoA $\theta$ is the unknown parameter to be estimated by the AP.
\subsubsection{Extended Target Model}
When the spatial extent of the target is large, the echo signals reflected by the target come from several scatterers in an extended region of space \cite{259642,9652071,4200705,8579200}. In this case, the point target model does not accurately reflect the behavior of distributed point-like scatterers, as the echo signals reflected by the extended target consist of multiple signal paths from scatterers with different angles. It is assumed that the AP has no prior knowledge about the distribution of the scatterers. As a result, the AP needs to estimate the complete target response matrix $\mathbf H$, such that the angles of scatterers can be recovered from the estimation of the target response matrix using techniques such as the MUSIC algorithm\cite{506612,schmidt1986multiple,32276}.

Based on the transmitted signal $\mathbf x(t)$ and the channel models, the signal impinged at the IRS is $\mathbf G \mathbf x(t)$. After the reflective beamforming at the IRS and target reflection, the echo signal impinged at the IRS becomes $\mathbf H \mathbf \Phi \mathbf G \mathbf x(t)$, where $\mathbf{\Phi} = \mathrm {diag}(\mathbf v)$ denotes the reflection matrix of the IRS. By further reflective beamforming at IRS and through the IRS-AP channel, the received echo signal at the AP through the whole AP-IRS-target-IRS-AP link at time $t\in\{1,..., T\}$ is
\begin{equation}\label{eq:echo_signal}
\mathbf y(t)=\mathbf{G}^\mathrm{T}\mathbf{\Phi}^\mathrm{T}\mathbf H \mathbf{\Phi}{\mathbf{G}}\mathbf x(t)+ \mathbf n(t),
\end{equation}
where $\mathbf n(t) \sim \mathcal{C N}(\mathbf{0}, \sigma_\text{R}^2\mathbf I_M)$ denotes the additive white Gaussian noise (AWGN) at the AP receiver.
We stack the transmitted signals, the received signals, and the noise over the radar dwell time as $\mathbf X =[\mathbf x(1),\ldots,\mathbf x(T)]$, $\mathbf Y =[\mathbf y(1),\ldots,\mathbf y(T)]$, and $\mathbf N =[\mathbf n(1),\ldots,\mathbf n(T)]$, respectively. Accordingly, we have
\begin{equation}\label{eq: Y_normal}
\mathbf Y=\mathbf{G}^\mathrm{T}\mathbf{\Phi}^\mathrm{T}\mathbf H \mathbf{\Phi}{\mathbf{G}}\mathbf X+ \mathbf N.
\end{equation}

Based on the received echo signal in \eqref{eq: Y_normal}, our objective is to estimate the target's DoA $\theta$ in the point target case, and to estimate the target response matrix $\mathbf H$ in the extended target case. It is assumed that the AP perfectly knows the CSI $\mathbf G$ of the AP-IRS link via proper channel estimation algorithms (see, e.g., \cite{9722893}). This assumption is practically valid, which is due to the fact that the AP and the IRS are deployed at fixed locations and thus their channels are slowly varying in practice. It is also assumed that the AP perfectly knows its transmitted signal $\mathbf X$ (and the associated sample coherence matrix $\mathbf R_x$) and the IRS's reflective beamforming vector $\mathbf{v}$ (or $\mathbf{\Phi}$), which can be optimized to enhance the sensing performance.

In the following, Section III obtains the closed-form CRB expressions for target parameters estimation. Sections IV and V minimize the CRB in \eqref{eq:CRB} or \eqref{eq:CRB_1} by jointly optimizing the sample coherence matrix $\mathbf R_x$ at the AP and the reflective beamforming vector $\mathbf{v}$ at the IRS for the point target case. Sections V minimize the CRB in \eqref{eq:CRB_extended} by optimizing the reflective beamforming vector $\mathbf{v}$ at the IRS for the extended target case.

\section{Estimation CRB}
This section analyzes the CRB performance in the IRS-enabled NLoS wireless sensing system for the point and extended target cases, respectively.

\subsection{Point Target Case}
We first derive CRB for estimating the DoA $\theta$ in the point target case. Let $\bm \xi=[\theta,  \tilde{\bm\alpha}^\mathrm{T}]^\mathrm{T} \in \mathbb{R}^{3\times 1}$ denote the vector of unknown parameters to be estimated, including the target's DoA $\theta$ and the complex-valued channel coefficient $\alpha$, where $\tilde{\bm\alpha}=[\mathrm{Re}\{\alpha\},\mathrm{Im}\{\alpha\}]^\mathrm{T}$.  We are particularly interested in characterizing the CRB for estimating the target's DoA $\theta$. This is due to the fact that it is difficult to extract the target information from the channel coefficient $\alpha$, as it depends on both the target's RCS and the distance-dependent path loss of the IRS-target-IRS link that are usually unknown.

First, we obtain the Fisher information matrix (FIM) for estimating $\bm \xi$ to facilitate the derivation of the CRB for DoA estimation. Substituting the target response matrix in \eqref{eq:channel_point} into \eqref{eq: Y_normal}, we have
\begin{equation}\label{eq: Y}
\mathbf Y = \alpha\mathbf B(\theta)\mathbf X+\mathbf N,
\end{equation}
where $\mathbf B(\theta)= \mathbf b(\theta)\mathbf b(\theta)^\mathrm{T}$ with $\mathbf b(\theta)= \mathbf{G}^\mathrm{T}\mathbf{\Phi}^\mathrm{T}\mathbf a(\theta)$. For notational convenience, in the sequel we drop $\theta$ in $\mathbf a(\theta)$, $\mathbf b(\theta)$, and $\mathbf B(\theta)$, and accordingly denote them as $\mathbf a$, $\mathbf b$, and $\mathbf B$, respectively. By vectorizing \eqref{eq: Y}, we have
\begin{equation}\label{eq:vec_data}
\tilde{\mathbf y}=\mathrm{vec}(\mathbf Y)=\tilde{\mathbf u} + \tilde{\mathbf n},
\end{equation}
where $\tilde{\mathbf u} = \alpha \mathrm{vec}(\mathbf B \mathbf X)$ and $\tilde{\mathbf n}=\mathrm{vec}(\mathbf N) \sim \mathcal{C N}(\mathbf{0}, \mathbf{R}_n)$ with $\mathbf R_n=\sigma_\text{R}^2\mathbf I_{MT}$.
Let $\tilde{\mathbf F}  \in \mathbb{R}^{3 \times 3}$ denote the FIM for estimating $\bm \xi$ based on \eqref{eq:vec_data}. Since $\tilde{\mathbf n}$ is AWGN, each element of $\tilde{\mathbf F}$ is given by \cite{kay1993fundamentals}
\begin{equation}\label{eq:FIM}
\begin{split}
\tilde{\mathbf F}_{i,j}=&\mathrm{tr}\left(\mathbf R_n^{-1}\frac{\partial \mathbf R_n}{\partial \bm \xi_i}\mathbf R_n^{-1}\frac{\partial \mathbf R_n}{\partial \bm \xi_j}\right)\\
&+2\mathrm{Re}\left\{\frac{\partial \tilde{\mathbf u}^\mathrm{H}}{\partial \bm \xi_i}\mathbf R_n^{-1}\frac{\partial \tilde{\mathbf u}}{\partial \bm \xi_j}\right\}, i,j\in\{1,2,3\}.
\end{split}
\end{equation}
Based on \eqref{eq:FIM}, the FIM $\tilde{\mathbf F}$ is partitioned as
\begin{equation}\label{eq:FIM_partitioned}
\tilde{\mathbf F}=
\begin{bmatrix}
\tilde{\mathbf{F}}_{\theta \theta} & \tilde{\mathbf{F}}_{\theta \tilde{\bm\alpha}}\\
\tilde{\mathbf{F}}^\mathrm{T}_{\theta \tilde{\bm\alpha}} & \tilde{\mathbf{F}}_{\tilde{\bm\alpha} \tilde{\bm\alpha}}
\end{bmatrix},
\end{equation}
where
\begin{equation}\label{eq:FIM_1}
\tilde{\mathbf F}_{\theta \theta}=\frac{2T|\alpha|^2}{\sigma_\text{R}^2}\text{tr}(\dot {\mathbf B}  \mathbf R_x \dot {\mathbf B}^\mathrm{H} ),
\end{equation}
\begin{equation}\label{eq:FIM_2}
\tilde{\mathbf F}_{\theta \tilde{\bm \alpha}}=\frac{2T}{\sigma_\text{R}^2}\mathrm{Re}\{\alpha^*\text{tr}( {\mathbf B}\mathbf R_x \dot {\mathbf B}^\mathrm{H} )[1,j]\},
\end{equation}
\begin{equation}\label{eq:FIM_3}
\tilde{\mathbf F}_{\tilde{\bm\alpha} \tilde{\bm\alpha}}=\frac{2T}{\sigma_\text{R}^2}\text{tr}( {\mathbf B} \mathbf R_x  {\mathbf B^\mathrm{H}} )\mathbf I_2,
\end{equation}
with $j = \sqrt{-1}$ and $\dot{\mathbf B} =\frac{\partial \mathbf B }{\partial \theta}$ denoting the partial derivative of $\mathbf B$ w.r.t. $\theta$. The derivation of the FIM $\tilde{\mathbf F}$ follows the standard procedure in \cite{bekkerman2006target}, see the details in Appendix A.

Next, we derive the CRB for estimating the DoA, which corresponds to the first diagonal element of $\tilde{\mathbf F}^{-1}$, i.e.,
\begin{equation}\label{eq:FIM_the}
\mathrm{CRB}(\theta) =[\tilde{\mathbf F}^{-1}]_{1,1} =[\tilde{\mathbf F}_{\theta \theta}-\tilde{\mathbf F}_{\theta \tilde{\bm\alpha}}\tilde{\mathbf F}_{\tilde{\bm\alpha} \tilde{\bm\alpha}}\tilde{\mathbf F}_{\theta \tilde{\bm\alpha}}^\mathrm{T}]^{-1}.
\end{equation}
Based on \eqref{eq:FIM_the} and \eqref{eq:FIM_partitioned}, we have the following lemma.
\begin{lemma} \label{lemma1}
The CRB for estimating the DoA $\theta$ is given by
\begin{equation}\label{eq:CRB}
\mathrm{CRB}(\theta)=\frac{\sigma_\text{R}^2}{2T|\alpha|^2\left(\mathrm{tr}(\dot {\mathbf B} \mathbf R_x \dot {\mathbf B}^\mathrm{H})-\frac{|\mathrm{tr}(\mathbf B \mathbf R_x  \dot {\mathbf B}^\mathrm{H})|^2}{\mathrm{tr}(\mathbf B \mathbf R_x \mathbf B^\mathrm{H})}\right)}.
\end{equation}
\end{lemma}

To gain more insight and facilitate the reflective beamforming design, we re-express $\mathrm{CRB}(\theta)$ in \eqref{eq:CRB} w.r.t. the reflective beamforming vector $\mathbf v$. Towards this end, we introduce $\mathbf A=\mathrm{diag}(\mathbf a)$, and accordingly have $\mathbf b=\mathbf{G}^\mathrm{T}\mathbf{\Phi}^\mathrm{T}\mathbf a = \mathbf{G}^\mathrm{T}\mathbf A\mathbf{v}$. Then, let $\dot{\mathbf b}$ denote the partial derivative of $\mathbf b$ w.r.t. $\theta$, where $\dot{\mathbf b}=j 2\pi \frac{d_\text{IRS}}{\lambda_\text{R}} \cos\theta\mathbf{G}^\mathrm{T}\mathbf{\Phi}^\mathrm{T}\mathbf D\mathbf a= j 2\pi \frac{d_\text{IRS}}{\lambda_\text{R}} \cos\theta\mathbf{G}^\mathrm{T}\mathbf A\mathbf D\mathbf{v}$ with $\mathbf D = \mathrm{diag}(0,1,...,N-1)$. As a result, 
\begin{equation}\label{eq:B}
\mathbf B  = \mathbf b \mathbf b ^\mathrm{T}= \mathbf{G}^\mathrm{T}\mathbf A\mathbf{v} \mathbf v^\mathrm{T}\mathbf A^\mathrm{T}\mathbf G,
\end{equation}
\begin{equation}\label{eq:B_dot}
\begin{split}
\dot {\mathbf B}  =& \dot {\mathbf b} \mathbf b^\mathrm{T} +\mathbf b \dot {\mathbf b}^\mathrm{T}\\
=&j 2\pi \frac{d_\text{IRS}}{\lambda_\text{R}} \cos\theta \mathbf{G}^\mathrm{T}\mathbf A(\mathbf D\mathbf{v} \mathbf v^\mathrm{T}+ \mathbf{v} \mathbf v^\mathrm{T}\mathbf D^\mathrm{T})\mathbf A^\mathrm{T}\mathbf G.
\end{split}
\end{equation}
Substituting \eqref{eq:B} and \eqref{eq:B_dot} into \eqref{eq:CRB}, we re-express $\mathrm{CRB}(\theta)$  w.r.t. $\mathbf v$ in \eqref{eq:CRB_1} at the top of next page, where $\mathbf R_1=\mathbf A^\mathrm{H}\mathbf{G}^*\mathbf{G}^\mathrm{T}\mathbf A$ and $\mathbf R_2=\mathbf A^\mathrm{H}  \mathbf{G}^* \mathbf R^*_x \mathbf{G}^\mathrm{T}\mathbf A$.
\newcounter{TempEqCnt}
\begin{figure*}[ht]
\begin{equation}\label{eq:CRB_1}
\mathrm{CRB}(\theta)=\frac{\sigma_\text{R}^2 \lambda_\text{R}^2}{ 8T|\alpha|^2\pi^2d_\text{IRS}^2\cos^2(\theta)\left(\mathbf v^\mathrm{H} \mathbf R_2 \mathbf v \left(\mathbf v^\mathrm{H} \mathbf D \mathbf R_1 \mathbf D\mathbf v-\frac{|\mathbf v^\mathrm{H} \mathbf D \mathbf R_1 \mathbf v|^2}{\mathbf v^\mathrm{H} \mathbf R_1 \mathbf v}\right) + \mathbf v^\mathrm{H} \mathbf R_1 \mathbf v \left(\mathbf v^\mathrm{H} \mathbf D \mathbf R_2 \mathbf D \mathbf v-\frac{|\mathbf v^\mathrm{H} \mathbf D \mathbf R_2 \mathbf v|^2}{\mathbf v^\mathrm{H} \mathbf R_2 \mathbf v}\right)\right)}
\end{equation}
\hrulefill
\end{figure*}

Based on $\mathrm{CRB}(\theta)$ in \eqref{eq:CRB_1}, we have the following proposition.
\begin{proposition} \label{proposition1}
If $\mathrm{rank}(\mathbf G)= 1$ (or equivalently there is only one single path between the AP and the IRS), then the FIM $\tilde{\mathbf F}$ in \eqref{eq:FIM} for estimating $\bm \xi$ is a singular matrix, and $\mathrm{CRB}(\theta) = \infty$. Otherwise, the FIM $\tilde{\mathbf F}$ is invertible, and $\mathrm{CRB}(\theta)$ is bounded.
\end{proposition}
\begin{IEEEproof}
See Appendix B.
\end{IEEEproof}

Proposition \ref{proposition1} shows that the target's DoA $\theta$ is estimable only when $\mathrm{rank}(\mathbf G)>1$ (i.e., the number of signal paths between the AP and the IRS is larger than one). The reasons are intuitively explained as follows.  When $\mathrm{rank}(\mathbf G)= 1$, the truncated SVD of $\mathbf G$ is expressed as $\mathbf G= \sigma_1 \mathbf s_1 \mathbf q_1^\mathrm{T}$, where $\mathbf s_1$ and $\mathbf q_1$ are the left and right dominant singular vectors, respectively, and $\sigma_1$ denotes the dominant singular value. Accordingly, the received echo signal in \eqref{eq:echo_signal} at the AP is given by
\begin{equation}\label{eq:echo_signal_G}
\mathbf y(t)= \alpha\sigma_1^2 \mathbf q_1 \underbrace{\mathbf s_1^\mathrm{T}\mathbf {\mathbf{\Phi}}^\mathrm{T}\mathbf a\mathbf a^\mathrm{T} \mathbf{\Phi}\mathbf s_1}_{\beta(\theta)} \underbrace{\mathbf q_1^\mathrm{T}\mathbf x(t)}_{\tilde x(t)} + \mathbf n(t),
\end{equation}
where $\beta(\theta) =  |\mathbf s_1^\mathrm{T}\mathbf {\mathbf{\Phi}}^\mathrm{T}\mathbf a|^2$ is the only scalar term related to $\theta$. Notice that in \eqref{eq:echo_signal_G}, the complex numbers $\alpha$ and $\beta(\theta)$ are coupled. Therefore,  from $\mathbf y(t)$, we can only recover one observation on $\alpha \beta(\theta)$, which is not sufficient to extract $\beta(\theta)$, thus making the DoA $\theta$ not estimable.

\subsection{Extended Target Case}
This subsection derives  the CRB for estimating the target response matrix $\mathbf H$. In this case, let $\bm \zeta=[\mathbf h_\text{R}^\mathrm{T}, \mathbf h_\text{I}^\mathrm{T}]^\mathrm{T} \in \mathbb{R}^{2N^2\times 1}$  denote the vector of unknown parameters to be estimated, where $\mathbf h_\text{R} = \mathrm{Re}(\mathbf h)$ and $\mathbf h_\text{I} = \mathrm{Im}(\mathbf h)$ are the real and imaginary parts of $\mathbf h = \mathrm{vec}(\mathbf H)$, respectively.

First, we obtain the FIM for estimating the target response matrix $\mathbf H$.
Similarly as in \eqref{eq:vec_data}, for the extended target case, the received echo signal is rewritten as
\begin{equation}\label{eq:vec_data_extended}
\hat{\mathbf y}=\mathrm{vec}(\mathbf Y)=\hat{\mathbf u} + \hat{\mathbf n},
\end{equation}
where $\hat{\mathbf u} = \mathrm{vec}(\mathbf{G}^\mathrm{T}\mathbf{\Phi}^\mathrm{T}\mathbf H \mathbf{\Phi}{\mathbf{G}}\mathbf X)=(\mathbf X^\mathrm{T}\mathbf{G}^\mathrm{T}\mathbf{\Phi}^\mathrm{T}\otimes \mathbf{G}^\mathrm{T}\mathbf{\Phi}^\mathrm{T})\mathbf h$ and $\hat{\mathbf n}=\mathrm{vec}(\mathbf N) \sim \mathcal{C N}(\mathbf{0}, \mathbf{R}_n)$ with $\mathbf R_n=\sigma_\text{R}^2\mathbf I_{MT}$.
Let $\hat{\mathbf F}  \in \mathbb{R}^{2N^2 \times 2N^2}$ denote the FIM for estimating $\bm \zeta$ based on \eqref{eq:vec_data_extended}. Similarly as for \eqref{eq:FIM}, each element of $\hat{\mathbf F}$ is given by
\begin{equation}\label{eq:FIM_extended}
\begin{split}
\hat{\mathbf F}_{i,j}=&\mathrm{tr}\left(\mathbf R_n^{-1}\frac{\partial \mathbf R_n}{\partial \bm \zeta_i}\mathbf R_n^{-1}\frac{\partial \mathbf R_n}{\partial \bm \zeta_j}\right)\\
&+2\mathrm{Re}\left\{\frac{\partial \hat{\mathbf u}^\mathrm{H}}{\partial \bm \zeta_i}\mathbf R_n^{-1}\frac{\partial \hat{\mathbf u}}{\partial \bm \zeta_j}\right\}, i,j\in\{1,\ldots,2N^2\}.
\end{split}
\end{equation}

Based on \eqref{eq:FIM_extended}, the FIM $\hat{\mathbf F}$ is partitioned as
\begin{equation}\label{eq:FIM_partitioned_extended}
\hat{\mathbf F}=
\begin{bmatrix}
\hat{\mathbf{F}}_{\mathbf h_\text{R} \mathbf h_\text{R}} & \hat{\mathbf{F}}_{\mathbf h_\text{R} \mathbf h_\text{I}}\\
\hat{\mathbf{F}}_{\mathbf h_\text{I} \mathbf h_\text{R}} & \hat{\mathbf{F}}_{\mathbf h_\text{I} \mathbf h_\text{I}}
\end{bmatrix},
\end{equation}
where
\begin{equation}\label{eq:FIM_1_extended}
\begin{split}
&\hat{\mathbf F}_{\mathbf h_\text{R} \mathbf h_\text{R}}=\hat{\mathbf{F}}_{\mathbf h_\text{I} \mathbf h_\text{I}}\\
=&\frac{2T}{\sigma_\text{R}^2}\mathrm{Re}\{(\mathbf{\Phi}^*\mathbf{G}^*\mathbf R_X^\mathrm{T}\mathbf{G}^\mathrm{T}\mathbf{\Phi}^\mathrm{T})\otimes (\mathbf{\Phi}^*\mathbf{G}^*\mathbf{G}^\mathrm{T}\mathbf{\Phi}^\mathrm{T})\},
\end{split}
\end{equation}
\begin{equation}\label{eq:FIM_2_extended}
\begin{split}
&\hat{\mathbf F}_{\mathbf h_\text{I} \mathbf h_\text{R}}=-\hat{\mathbf{F}}_{\mathbf h_\text{R} \mathbf h_\text{I}}\\
=&\frac{2T}{\sigma_\text{R}^2}\mathrm{Im}\{(\mathbf{\Phi}^*\mathbf{G}^*\mathbf R_X^\mathrm{T}\mathbf{G}^\mathrm{T}\mathbf{\Phi}^\mathrm{T})\otimes (\mathbf{\Phi}^*\mathbf{G}^*\mathbf{G}^\mathrm{T}\mathbf{\Phi}^\mathrm{T})\},
\end{split}
\end{equation}
see the details in Appendix C. Based on \eqref{eq:FIM_partitioned_extended}, the CRB for estimating $\mathbf H$ is obtained in the following lemma.
\begin{lemma} \label{lemma2}
The CRB for estimating the target response matrix $\mathbf H$ is given by
\begin{equation}\label{eq:CRB_extended}
\mathrm{CRB}(\mathbf H)
=\frac{\sigma_\text{R}^2}{T}\mathrm{tr}((\mathbf G\mathbf R_x\mathbf G^\mathrm{H})^{-1}) \mathrm{tr}((\mathbf G\mathbf G^\mathrm{H})^{-1}).
\end{equation}
\end{lemma}
\begin{IEEEproof}
It follows from \eqref{eq:FIM_partitioned_extended} that
\begin{equation}
\begin{split}
\mathrm{CRB}(\mathbf H)=& \mathrm{CRB}(\bm \zeta) =\mathrm{tr}(\hat{\mathbf F}^{-1})\\
=&\frac{\sigma_\text{R}^2}{T}\mathrm{tr}((\mathbf{\Phi}\mathbf G\mathbf R_x\mathbf G^\mathrm{H}\mathbf{\Phi}^\mathrm{H})^{-1}) \mathrm{tr}((\mathbf{\Phi}\mathbf G\mathbf G^\mathrm{H}\mathbf{\Phi}^\mathrm{H})^{-1})\\
=&\frac{\sigma_\text{R}^2}{T}\mathrm{tr}((\mathbf G\mathbf R_x\mathbf G^\mathrm{H})^{-1}) \mathrm{tr}((\mathbf G\mathbf G^\mathrm{H})^{-1}),
\end{split}
\end{equation}
completing the proof.
\end{IEEEproof}

It is observed from Lemma \ref{lemma2} that $\mathrm{CRB}(\mathbf H)$ in the extended target case only depends on the transmit beamformers or the sample coherence matrix $\mathbf R_x$, regardless of the reflective beamformer $\mathbf \Phi$. Based on FIM $\hat{\mathbf F}$ in \eqref{eq:FIM_partitioned_extended} and $\mathrm{CRB}(\mathbf H)$ in \eqref{eq:CRB_extended}, we have the following proposition.
\begin{proposition}\label{proposition2}
If $\mathrm{rank}(\mathbf G)< N$, then the FIM $\hat{\mathbf F}$ in \eqref{eq:FIM_partitioned_extended} for estimating $\mathbf H$ is a singular matrix, and $\mathrm{CRB}(\mathbf H) = \infty$. Otherwise, the FIM $\hat{\mathbf F}$ is invertible, and $\mathrm{CRB}(\mathbf H)$ is bounded.
\end{proposition}

Proposition \ref{proposition2} shows that the target response matrix $\mathbf H$ is estimable only when $\mathrm{rank}(\mathbf G) = N$.

\section{Joint Beamforming Optimization for CRB Minimization with Point Target}
In this section, we propose to jointly optimize the transmit beamforming at the AP and the reflective beamforming at the IRS to minimize the CRB for estimating the DoA in \eqref{eq:CRB} or \eqref{eq:CRB_1} in the point target case, subject to the maximum transmit power constraint at the AP. Note that in order to implement the joint beamforming design, we assume that the AP roughly knows the information of $\theta$, which is practically valid for a target tracking scenario. Also note that the AP does not need to know the channel coefficient $\alpha$, as it is independent of the joint beamforming design. The CRB minimization problem is formulated as
\begin{subequations}
\begin{align} \notag
  \text{(P1)}:  \ \  \min_{\mathbf R_x, \mathbf{v}} \  \  &\mathrm{CRB}(\theta) \\ \label{eq:power}
  \text{s.t.}  \quad   &\mathrm{tr}(\mathbf R_x) \leq P_0\\\label{eq:semi}
   &\mathbf R_x \succeq \mathbf{0}\\ \label{eq:phase_Phi}
   &\mathbf{\Phi}=\mathrm{diag}(\mathbf v)\\\label{eq:phase_1}
  & |\mathbf v_n|=1, \forall n\in \{1,...,N\},
\end{align}
\end{subequations}
where $P_0$ is the maximum transmit power at the AP. Problem (P1) is non-convex due to the non-convexity of the objective function and the unit-modulus constraint in \eqref{eq:phase_1}.

To solve the non-convex problem (P1), we propose an efficient algorithm based on alternating optimization, in which the transmit beamformers $\mathbf R_x$ at the AP and the reflective beamformer $\mathbf v$ at the IRS are optimized in an alternating manner. 

\subsection{Transmit Beamforming Optimization}
First, we optimize the transmit beamformers $\mathbf R_x$ in problem (P1) under any given reflective beamformer $\mathbf v$, in which the CRB formula in \eqref{eq:CRB} is used. In this case, minimizing $\mathrm{CRB}(\theta)$ is equivalent to maximizing $\mathrm{tr}(\dot {\mathbf B} \mathbf R_x \dot {\mathbf B}^\mathrm{H})-\frac{|\mathrm{tr}(\mathbf B\mathbf R_x  \dot {\mathbf B}^\mathrm{H})|^2}{\mathrm{tr}(\mathbf B \mathbf R_x \mathbf B^\mathrm{H})}$. As a result, the transmit beamforming optimization problem is formulated as
\begin{subequations}
\begin{align} \notag
  \text{(P2)}: \ \ \max_{\mathbf R_x} & \  \  \mathrm{tr}(\dot {\mathbf B} \mathbf R_x \dot {\mathbf B}^\mathrm{H})-\frac{|\mathrm{tr}(\mathbf B\mathbf R_x  \dot {\mathbf B}^\mathrm{H})|^2}{\mathrm{tr}(\mathbf B \mathbf R_x \mathbf B^\mathrm{H})} \\\notag
  \text{s.t.} &  \quad \eqref{eq:power}~\text{and}  ~\eqref{eq:semi}.
\end{align}
\end{subequations}
By introducing an auxiliary variable $t$, problem (P2) is equivalently re-expressed as
\setcounter{equation}{27}
\begin{subequations}
\begin{align} \notag
  \text{(P2.1)}: \ \ \max_{\mathbf R_x,t}&  \  \  t \\  \label{eq:Schur's complement}
  \text{s.t.}&   \ \ \mathrm{tr}(\dot {\mathbf B} \mathbf R_x \dot {\mathbf B}^\mathrm{H})-\frac{|\mathrm{tr}(\mathbf B\mathbf R_x  \dot {\mathbf B}^\mathrm{H})|^2}{\mathrm{tr}(\mathbf B \mathbf R_x \mathbf B^\mathrm{H})} \ge t\\ \notag
  & \quad  \eqref{eq:power}~\text{and}~ \eqref{eq:semi}.
\end{align}
\end{subequations}
Using the Schur complement\cite{zhang2006schur}, the constraint in \eqref{eq:Schur's complement} is equivalently transformed into the following convex semi-definite constraint:
\begin{equation}\label{eq:SC}
\left[\begin{array}{cc}
      \mathrm{tr}(\dot {\mathbf B} \mathbf R_x \dot {\mathbf B}^\mathrm{H})-t & \mathrm{tr}(\mathbf B \mathbf R_x  \dot {\mathbf B}^\mathrm{H})  \\
      \mathrm{tr}(\dot {\mathbf B} \mathbf R_x \mathbf B^\mathrm{H} ) &  \mathrm{tr}(\mathbf B \mathbf R_x\mathbf B^\mathrm{H})
\end{array}\right] \succeq \mathbf{0}.
\end{equation}
Accordingly, problem (P2.1) is equivalent to the following semi-definite program (SDP), which can be optimally solved by convex solvers such as CVX \cite{cvx}:
\begin{subequations}
\begin{align} \notag
  \text{(P2.2)}: \ \ & \max_{\mathbf R_x,t}  \  \  t \\  \notag
  \text{s.t.}&   \ \ \eqref{eq:power}, ~ \eqref{eq:semi}, ~ \text{and}~ \eqref{eq:SC}.
\end{align}
\end{subequations}

\subsection{Reflective Beamforming Optimization}
Next, we optimize the reflecting beamformer $\mathbf v$ in problem (P1) under any given  transmit beamformers $\mathbf R_x$, in which the CRB formula in \eqref{eq:CRB_1} is used. In this case, the reflective beamforming optimization problem is formulated as
\begin{subequations}
\begin{align} \notag
  \text{(P3)}:  \ \  \max_{\mathbf{v}}  \  \  &\left(\mathbf v^\mathrm{H} \mathbf R_2 \mathbf v \left(\mathbf v^\mathrm{H} \mathbf D \mathbf R_1 \mathbf D\mathbf v-\frac{|\mathbf v^\mathrm{H} \mathbf D \mathbf R_1 \mathbf v|^2}{\mathbf v^\mathrm{H} \mathbf R_1 \mathbf v}\right)  \right. \\ \notag
  &+ \left. \mathbf v^\mathrm{H} \mathbf R_1 \mathbf v \left(\mathbf v^\mathrm{H} \mathbf D \mathbf R_2 \mathbf D \mathbf v-\frac{|\mathbf v^\mathrm{H} \mathbf D \mathbf R_2 \mathbf v|^2}{\mathbf v^\mathrm{H} \mathbf R_2 \mathbf v}\right)\right) \\\notag
  \text{s.t.}  \ \  & \ \ \eqref{eq:phase_1},
\end{align}
\end{subequations}
which is still non-convex due to the non-concavity of the objective function and the unit-modulus constraint in \eqref{eq:phase_1}. To resolve this issue, we first deal with constraint \eqref{eq:phase_1} based on SDR, and then use SCA to approximate the relaxed problem.

Define $\mathbf{V}=\mathbf{v} \mathbf{v}^\mathrm {H}$ with $\mathbf{V} \succeq \mathbf{0}$ and $\mathrm {rank}(\mathbf{V})=1$. Based on \eqref{eq:phase_1}, $\mathbf V_{n,n}=1,\forall n\in \{1,2,..., N\}$. Also, we have  $\mathbf v^\mathrm{H} \mathbf R_1 \mathbf v = \mathrm{tr}(\mathbf R_1 \mathbf V)$, $\mathbf v^\mathrm{H} \mathbf R_2 \mathbf v = \mathrm{tr}(\mathbf R_2 \mathbf V)$, $\mathbf v^\mathrm{H} \mathbf D \mathbf R_1 \mathbf v = \mathrm{tr}(\mathbf D \mathbf R_1 \mathbf V)$, $\mathbf v^\mathrm{H} \mathbf D \mathbf R_2 \mathbf v = \mathrm{tr}(\mathbf D \mathbf R_2 \mathbf V)$, $\mathbf v^\mathrm{H} \mathbf D \mathbf R_1 \mathbf D\mathbf v=\mathrm{tr}(\mathbf D \mathbf R_1 \mathbf D\mathbf V)$, and $\mathbf v^\mathrm{H} \mathbf D \mathbf R_2 \mathbf D\mathbf v=\mathrm{tr}(\mathbf D \mathbf R_2 \mathbf D\mathbf V)$. By substituting $\mathbf{V}=\mathbf{v} \mathbf{v}^\mathrm {H}$ and introducing two auxiliary variables $t_1$ and $t_2$, problem (P3) is equivalently re-expressed as
\setcounter{equation}{29}
\begin{subequations}
\begin{align} \notag
  \text{(P3.1)}:  \max_{\mathbf V, t_1, t_2} \  \  &f_1(\mathbf V,t_1,t_2)+f_2(\mathbf V,t_1,t_2)\\\label{eq:p3_st_1}
  \text{s.t.}  \ \  &\mathbf V_{n,n}=1,\forall n\in \{1,2,\cdots, N\}\\\label{eq:p3_st_2}
  \qquad & \mathbf{V} \succeq \mathbf{0}\\ \label{eq:rank_1}
  \qquad & \mathrm {rank}(\mathbf{V})=1 \\ \label{eq:p3.3st_1}
  \qquad &t_1\ge \frac{|\mathrm{tr}(\mathbf D\mathbf R_1\mathbf{V})|^2}{\mathrm{tr}(\mathbf R_1\mathbf{V})}\\\label{eq:p3.3st_2}
  \qquad &t_2\ge \frac{|\mathrm{tr}(\mathbf D\mathbf R_2\mathbf{V})|^2}{\mathrm{tr}(\mathbf R_2\mathbf{V})},
\end{align}
\end{subequations}
where
\begin{equation}
\begin{split}
&f_1(\mathbf V,t_1,t_2)\\
=&\frac{1}{4}\mathrm{tr}^2((\mathbf R_2+\mathbf D\mathbf R_1\mathbf D)\mathbf{V})+\frac{1}{4}(\mathrm{tr}(\mathbf R_2 \mathbf V-t_1)^2\\
  &+\frac{1}{4}\mathrm{tr}^2((\mathbf R_1+\mathbf D\mathbf R_2\mathbf D)\mathbf{V})+\frac{1}{4}(\mathrm{tr}(\mathbf R_1 \mathbf V-t_2)^2
\end{split}
\end{equation}
 and
\begin{equation}
\begin{split}
&f_2(\mathbf V,t_1,t_2)\\
=&-\frac{1}{4}\mathrm{tr}^2((\mathbf R_2-\mathbf D\mathbf R_1\mathbf D)\mathbf{V})-\frac{1}{4}(\mathrm{tr}(\mathbf R_2 \mathbf V+t_1)^2\\
  &-\frac{1}{4}\mathrm{tr}^2((\mathbf R_1-\mathbf D\mathbf R_2\mathbf D)\mathbf{V})-\frac{1}{4}(\mathrm{tr}(\mathbf R_1 \mathbf V+t_2)^2.
\end{split}
\end{equation}
Here, $f_1(\mathbf V,t_1,t_2)$ and  $f_2(\mathbf V,t_1,t_2)$ are convex and concave functions, respectively.
Using the Schur complement\cite{zhang2006schur}, the constraints in \eqref{eq:p3.3st_1} and \eqref{eq:p3.3st_2} are equivalently transformed into the following convex semi-definite constraints:
\begin{equation}\label{eq:sc_2}
   \left[\begin{array}{cc}
      t_1 & \mathrm{tr}(\mathbf D\mathbf R_1 \mathbf V)  \\
      \mathrm{tr}(\mathbf V^\mathrm{H}\mathbf R_1^\mathrm{H} \mathbf D^\mathrm{H}) &  \mathrm{tr}(\mathbf R_1 \mathbf V)
\end{array}\right] \succeq \mathbf{0},
\end{equation}
\begin{equation}\label{eq:sc_3}
\left[\begin{array}{cc}
      t_2 & \mathrm{tr}(\mathbf D\mathbf R_2 \mathbf V)  \\
      \mathrm{tr}(\mathbf V^\mathrm{H}\mathbf R_2^\mathrm{H} \mathbf D^\mathrm{H}) &  \mathrm{tr}(\mathbf R_2 \mathbf V)
\end{array}\right] \succeq \mathbf{0}.
\end{equation}

Problem (P3.1) is then equivalent to the following problem:
\begin{subequations}
\begin{align} \notag
  \text{(P3.2)}:  \max_{\mathbf V, t_1, t_2} & \  \  f_1(\mathbf V,t_1,t_2)+f_2(\mathbf V,t_1,t_2)\\\notag
  \text{s.t.}  \ \  &\eqref{eq:p3_st_1},~\eqref{eq:p3_st_2},~\eqref{eq:rank_1},~\eqref{eq:sc_2},~\text{and}~\eqref{eq:sc_3}.
\end{align}
\end{subequations}
By dropping the rank-one constraint in \eqref{eq:rank_1}, we get the relaxed version of problem (P3.2) without constraint \eqref{eq:rank_1}, denoted by problem (P3.3):
\begin{subequations}
\begin{align} \notag
  \text{(P3.3)}:  \max_{\mathbf V, t_1, t_2} & \  \  f_1(\mathbf V,t_1,t_2)+f_2(\mathbf V,t_1,t_2)\\\notag
  \text{s.t.}  \ \  &\eqref{eq:p3_st_1},~\eqref{eq:p3_st_2},~\eqref{eq:sc_2},~\text{and}~\eqref{eq:sc_3}.
\end{align}
\end{subequations}
Note that in problem (P3.3), all constraints are convex, but $f_1(\mathbf V,t_1,t_2)$ in the objective function is non-concave, thus making problem (P3.3) non-convex.

We use SCA to deal with the non-concave function $f_1(\mathbf V,t_1,t_2)$ for solving the non-convex problem (P3.3), which approximates (P3.3) as a series of convex problems. The SCA-based solution to problem (P3.3) is implemented in an iterative manner as follows. Consider inner iteration $r \ge 1$, in which the local point is denoted by $\mathbf V^{(r)}$, $t_1^{(r)}$, and $t_2^{(r)}$. Based on the local point, we obtain a global linear lower bound function $\hat f_1^{(r)}(\mathbf V,t_1,t_2)$ for the convex function $f_1(\mathbf V,t_1,t_2)$ in  (P3.3), using its first-order Taylor expansion, i.e.,
\setcounter{equation}{34}
\begin{equation}
\begin{split}
&f_1(\mathbf V,t_1,t_2)\\
\ge& f_1(\mathbf V^{(r)},t_1^{(r)},t_2^{(r)})+\frac{1}{2}\mathrm{tr}((\mathbf R_2+\mathbf D\mathbf R_1\mathbf D)\mathbf{V}^{(r)})\\
&\mathrm{tr}((\mathbf R_2+\mathbf D\mathbf R_1\mathbf D)(\mathbf{V}-\mathbf{V}^{(r)}))+\frac{1}{2}\mathrm{tr}((\mathbf R_1\\
&+\mathbf D\mathbf R_2\mathbf D)\mathbf{V}^{(r)})\mathrm{tr}((\mathbf R_1+\mathbf D\mathbf R_2\mathbf D)(\mathbf{V}\!-\!\mathbf{V}^{(r)}))\\
&+\frac{1}{2}t_1^{(r)}(t_1-t_1^{(r)})+\frac{1}{2}t_2^{(r)}(t_2-t_2^{(r)})\\
&+\frac{1}{2}\mathrm{tr}(\mathbf R_2\mathbf{V}^{(r)})\mathrm{tr}(\mathbf R_2(\mathbf V-\mathbf{V}^{(r)}))\\
&+\frac{1}{2}\mathrm{tr}(\mathbf R_1\mathbf{V}^{(r)})\mathrm{tr}(\mathbf R_1(\mathbf V-\mathbf{V}^{(r)}))\\
\triangleq &\hat f_1^{(r)}(\mathbf V,t_1,t_2).
\end{split}
\end{equation}

Replacing $f_1(\mathbf V,t_1,t_2)$ by $\hat f_1^{(r)}(\mathbf V,t_1,t_2)$, problem (P3.3) is approximated as the following convex form in iteration $r$:
\begin{subequations}
\begin{align}\notag
 \text{(P3.4.}r\text{)}: \max_{\mathbf V,t_1,t_2}&  \quad  \hat f_1^{(r)}(\mathbf V,t_1,t_2)+f_2(\mathbf V,t_1,t_2)\\\notag
  \text{s.t.} & \quad  \eqref{eq:p3_st_1},~\eqref{eq:p3_st_2},~ \eqref{eq:sc_2},~\text{and}~ \eqref{eq:sc_3},
\end{align}
\end{subequations}
which can be optimally solved  by convex solvers such as CVX\cite{cvx}. Let $\hat{\mathbf V}^{\star}$, $\hat t_1^{\star}$, and $\hat t_2^{\star}$ denote the optimal solution to problem (P3.4.$r$), which is then updated to be the local point $\mathbf V^{(r+1)}$, $t_1^{(r+1)}$, and $t_2^{(r+1)}$ for the next inner iteration. Since $\hat f_1^{(r)}(\mathbf V,t_1,t_2)$ serves as a lower bound of $f_1(\mathbf V,t_1,t_2)$, it is ensured that $f_1(\mathbf V^{(r+1)}, t_1^{(r+1)}, t_2^{(r+1)}) + f_2(\mathbf V^{(r+1)}, t_1^{(r+1)}, t_2^{(r+1)})\ge f_1(\mathbf V^{(r)}, t_1^{(r)}, t_2^{(r)})+ f_2(\mathbf V^{(r)}, t_1^{(r)}, t_2^{(r)})$, i.e., the inner iteration leads to a non-decreasing objective value for problem (P3.3). Therefore, the convergence of SCA for solving problem (P3.3) is ensured. Let $\tilde{\mathbf V}$, $\tilde t_1$, and $\tilde t_2$ denote the obtained solution to problem (P3.3) based on SCA, where $\mathrm{rank}(\tilde {\mathbf V}) >1$ may hold in general.

Finally, we construct an approximate rank-one solution of $\mathbf V$ to problem (P3.2) or (P3) by using Gaussian randomization. Specifically, we first generate a number of random realizations $\mathbf z \sim \mathcal{CN}(\mathbf{0},\tilde{\mathbf V})$, and construct a set of candidate feasible solutions as
\setcounter{equation}{35}
\begin{equation}\label{eq:Gaussion}
\mathbf{v}=e^{j\mathrm {arg}(\mathbf z)}.
\end{equation}
We then choose the best $\mathbf v$ that achieves the maximum objective value of problem (P3.2) or (P3). Note that the Gaussian randomization should be implemented sufficiently many times to ensure that the objective value increases at each outer iteration of alternating optimization.

\subsection{Complete Algorithm of Alternating Optimization}
By combining Sections IV-A and IV-B, the alternating optimization based algorithm for solving problem (P1) is complete, which is summarized as Algorithm 1 in Table I. In each outer iteration, we first solve problem (P2) to update the transmit beamformers $\mathbf R_x$ with the updated reflective beamformer $\mathbf v$ in the previous round, and then solve problem (P3) to update $\mathbf v$ with the updated $\mathbf R_x$. Note that in each outer iteration of alternating optimization, problem (P2) is optimally solved, thus leading to a non-increasing CRB value. Also notice that with sufficient number of Gaussian randomizations, the obtained solution to problem (P3) is also ensured to result in a monotonically non-increasing CRB. As a result, the convergence of the proposed alternating optimization based algorithm for solving problem (P1) is ensured.

\begin{table}
\centering{
\caption{Algorithm 1 for Joint Transmit and Reflective Beamforming Optimization with Point Target\label{tab:table1}}}
    \begin{algorithm}[H]
        \begin{enumerate}[a)]
            \item Set outer iteration index $l= 1$ and initialize the reflective beamformer with random phase shifts as $\mathbf{v}^{(l)}$ 
            \item \textbf{Repeat}: \begin{enumerate}[1)]
            				\item Under given reflective beamformer $\mathbf{v}^{(l)}$, solve problem (P2.2) to obtain the optimal solution as $\mathbf R_x^{(l)}$
            				\item Set inner iteration index  $r=1$, $\mathbf{V}^{(r)}=\mathbf{v}^{(l)}(\mathbf{v}^{(l)})^\mathrm{H}$, $t_1^{(r)}=\frac{|\mathrm{tr}(\mathbf D\mathbf R_1\mathbf{V}^{(r)})|^2}{\mathrm{tr}(\mathbf R_1\mathbf{V}^{(r)})}$, and $t_2^{(r)}=\frac{|\mathrm{tr}(\mathbf D\mathbf R_2\mathbf{V}^{(r)})|^2}{\mathrm{tr}(\mathbf R_2\mathbf{V}^{(r)})}$
            				\item \textbf{Repeat}: \begin{enumerate}[i)]
            						\item Construct function $\hat f_1^{(r)}(\mathbf V,t_1,t_2)$ using $\mathbf{V}^{(r)}$, $t_1^{(r)}$, and $t_2^{(r)}$
            						\item Solve problem (P3.3) under given $\mathbf R_x^{(l)}$ to obtain the optimal solution as $\mathbf{V}^{(r+1)}$, $t_1^{(r+1)}$, and $t_2^{(r+1)}$
            						\item Update $r=r+1$
            				\end{enumerate}
            				\item \textbf{Until} the convergence criterion is met or the maximum number of inner iterations is reached
            				\item Reconstruct an approximate rank-one solution $\mathbf{v}^{(l+1)}$ via \eqref{eq:Gaussion} based on the Gaussian randomization
            				\item Update $l=l+1$
            				\end{enumerate}
            				\item \textbf{Until} the convergence criterion is met or the maximum number of outer iterations is reached
            			
        \end{enumerate}
\centering{\caption{}}
\end{algorithm}
\end{table}

\section{Transmit Beamforming Optimization for CRB Minimization with Extended Target}
In this section, we optimize the transmit beamforming at the AP to minimize the CRB for estimating the target response matrix in \eqref{eq:CRB_extended} in the extended target case, subject to the maximum transmit power constraint at the AP. Based on Proposition $2$, we focus on the scenario with  $\mathrm{rank}(\mathbf G) = N$ and $M\ge N$, in order for $\mathrm{CRB}(\mathbf H)$ to be bounded or $\mathbf H$ to be estimable. Based on \eqref{eq:CRB_extended}, minimizing $\mathrm{CRB}(\mathbf H)$ is equivalent to minimizing $\mathrm{tr}((\mathbf G\mathbf R_x\mathbf G^\mathrm{H})^{-1})$.
As a result,  the CRB minimization problem for the extended target case is formulated as
\begin{subequations}
\begin{align} \notag
  \text{(P4)}:  \ \  \min_{\mathbf R_x} \  \  &\mathrm{tr}((\mathbf G\mathbf R_x\mathbf G^\mathrm{H})^{-1}) \\ \label{eq:P_constraint_energy}
  \text{s.t.}  \quad   &\mathrm{tr}(\mathbf R_x) \leq P_0\\ \label{eq:P_constraint_semi}
   &\mathbf R_x \succeq \mathbf{0}.
\end{align}
\end{subequations}
We obtain the optimal solution to the convex problem (P4) in closed-form in the following.

Let the SVD of $\mathbf G$ be denoted by $\mathbf G = \hat{\mathbf S} \hat{\mathbf \Sigma} \hat{\mathbf Q}^\mathrm{H}$, where $\hat{\mathbf S}\hat{\mathbf S}^\mathrm{H}=\hat{\mathbf S}^\mathrm{H}\hat{\mathbf S}=\mathbf I_N$, $\hat{\mathbf Q}\hat{\mathbf Q}^\mathrm{H}=\hat{\mathbf Q}^\mathrm{H}\hat{\mathbf Q}=\mathbf I_M$, and $\hat{\mathbf \Sigma} = [\hat{\mathbf \Sigma}_1, \mathbf 0] \in \mathbb{R}^{N \times M}$  with $\hat{\mathbf \Sigma}_1 = \mathrm{diag}(\hat{\sigma}_1, ..., \hat{\sigma}_N)$ and $\hat{\sigma}_1 \ge ... \ge \hat{\sigma}_N > 0$.
Then we have
\begin{equation}\label{eq:CRB_extended_ref}
\begin{split}
\mathrm{tr}((\mathbf G\mathbf R_x\mathbf G^\mathrm{H})^{-1}) =& \mathrm{tr}((\hat{\mathbf S} \hat{\mathbf \Sigma} \hat{\mathbf Q}^\mathrm{H}\mathbf R_x\hat{\mathbf Q} \hat{\mathbf \Sigma}^\mathrm{H} \hat{\mathbf S}^\mathrm{H})^{-1})\\
=&\mathrm{tr}((\hat{\mathbf \Sigma}^\mathrm{H} \hat{\mathbf \Sigma}\hat{\mathbf R}_x  )^{-1}),
\end{split}
\end{equation}
where
\begin{equation}\label{eq:R_X_Q}
\hat{\mathbf R}_x= \hat{\mathbf Q}^\mathrm{H}\mathbf R_x\hat{\mathbf Q}
\end{equation}
with $\hat{\mathbf R}_x \in \mathbb{C}^{M \times M}$ and $\hat{\mathbf R}_x \succeq \mathbf{0}$. Based on \eqref{eq:R_X_Q}, the power constraint in \eqref{eq:P_constraint_energy} is rewritten as
\begin{equation}\label{eq:power_new}
\mathrm{tr}(\mathbf{R_x})=\mathrm{tr}( \hat{\mathbf Q}^\mathrm{H}\mathbf R_x\hat{\mathbf Q})=\mathrm{tr}(\hat{\mathbf R}_x) \le P_0.
\end{equation}
Furthermore, define
\begin{equation}\label{eq:R_pati}
{\hat{\mathbf R}_{x}}=
\left[
  \begin{array}{cc}
    \hat{\mathbf R}_{x,1} & \hat{\mathbf R}_{x,2}\\
    \hat{\mathbf R}_{x,2}^\mathrm{H}  & \hat{\mathbf R}_{x,3}
  \end{array}
\right],
\end{equation}
where $\hat{\mathbf R}_{x,1}\in \mathbb{C}^{N \times N}$, $\hat{\mathbf R}_{x,2}\in \mathbb{C}^{N \times (M-N)}$, and $\hat{\mathbf R}_{x,3}\in \mathbb{C}^{(M-N) \times (M-N)}$. By substituting \eqref{eq:R_pati} into \eqref{eq:CRB_extended_ref}, we have 
\begin{equation}\label{eq:CRLB_extended_R_1}
\mathrm{tr}((\mathbf G\mathbf R_x\mathbf G^\mathrm{H})^{-1})= \mathrm{tr}((\hat{\mathbf \Sigma}_1^2 \hat{\mathbf R}_{x,1}  )^{-1}).
\end{equation}

Based on \eqref{eq:R_X_Q}, \eqref{eq:power_new}, and \eqref{eq:R_pati}, (P4) is reformulated as
\begin{subequations}
\begin{align} \notag
    \text{(P4.1)}: \min_{\hat{\mathbf R}_{x}\in \mathbb{C}^{N \times N}} \  \  &\mathrm{tr}((\hat{\mathbf \Sigma}_1^2 \hat{\mathbf R}_{x,1} )^{-1}) \\ \label{eq:power_P_4_1}
  \text{s.t.}  \quad   &\mathrm{tr}(\hat{\mathbf R}_{x,1}) + \mathrm{tr}(\hat{\mathbf R}_{x,3})\leq P_0\\ \label{eq:semi_hat_R}
   &\hat{\mathbf R}_{x} \succeq \mathbf{0}\\ \notag
  & \eqref{eq:R_pati}.
\end{align}
\end{subequations}

We next rely on the following lemma \cite{1327814,4194775}.
\begin{lemma}\label{lemma_J}
Let $\mathbf J \in \mathbb{C}^{N \times N}$ be a positive-definite Hermitian matrix with the $(i,i)\text{-th}$ entry $J_{i,i}$. Then 
\begin{equation}
\mathrm{tr}(\mathbf J^{-1})\ge \sum_{i=1}^{N}\frac{1}{ J_{i,i}},
\end{equation}
where the inequality is met with equality if and only if $\mathbf J$ is diagonal.
\end{lemma}

Based on Lemma \ref{lemma_J}, we have the following proposition.
\begin{proposition}\label{proposition_zero}
The optimality of problem (P4.1) is attained when $\hat{\mathbf R}_{x,2}$ and $\hat{\mathbf R}_{x,3}$ are all zero matrices, and $\hat{\mathbf R}_{x,1}$ is diagonal, i.e., $\hat{\mathbf R}_{x,1} = \mathrm{diag}(p_{1},...,p_{N})$, where $p_i \ge 0, \forall i \in \{1,...,N\}$.
\end{proposition}
\begin{IEEEproof}
First, assume that $\hat{\mathbf R}_{x} = \left[  \begin{array}{cc}
    \hat{\mathbf R}_{x,1} & \hat{\mathbf R}_{x,2}\\
    {\hat{\mathbf R}}_{x,2}^\mathrm{H}  & \hat{\mathbf R}_{x,3}
  \end{array}\right]
$ is optimal for problem (P4.1), where $\hat{\mathbf R}_{x,2}$ and $\hat{\mathbf R}_{x,3}$ are non-zero. Then, we reconstruct an alternative solution $\hat{\mathbf R}'_{x} = \left[  \begin{array}{cc}
    \hat{\mathbf R}'_{x,1} & \hat{\mathbf R}'_{x,2}\\
    {\hat{\mathbf R}'}_{x,2}^\mathrm{H}  & \hat{\mathbf R}'_{x,3}
  \end{array}\right]
$, where $\hat{\mathbf R}'_{x,1}=\frac{P_0}{\mathrm{tr}(\hat{\mathbf R}_{x,1})}\hat{\mathbf R}_{x,1}$, $\hat{\mathbf R}'_{x,2}=\mathbf 0$, and $\hat{\mathbf R}'_{x,3}=\mathbf 0$. It is easy to show that the alternative solution $\hat{\mathbf R}'_{x}$ satisfies the constraints in \eqref{eq:R_pati}, \eqref{eq:power_P_4_1}, and \eqref{eq:semi_hat_R}, and achieves a lower objective value for (P4.1) than that by $\hat{\mathbf R}_{x}$. Therefore, the presumption is not true, and the optimal solution of $\hat{\mathbf R}_{x,2}$ and $\hat{\mathbf R}_{x,3}$ should be both zero matrices. 

Next, suppose that $\hat{\mathbf R}_{x,1}$ is not diagonal. Then, we can construct an alternative solution as $\hat{\mathbf R}''_{x,1} = \mathrm{diag}(p_{1},...,p_{N})$, where $p_{i}$ is the $i$-th diagonal element of $\hat{\mathbf R}_{x,1}$, $i \in\{1,...,N\}$. Based on Lemma \ref{lemma_J}, it follows that the objective value achieved by $\hat{\mathbf R}''_{x,1}$ is smaller than that by $\hat{\mathbf R}_{x,1}$. Therefore, the presumption is not true. As a result, at the optimal solution of problem (P4.1), $\hat{\mathbf R}_{x,1}$ should be diagonal. This thus completes the proof.
\end{IEEEproof}

Based on Proposition \ref{proposition_zero}, the CRB minimization problem in (P4.1) is reformulated as
\begin{subequations}
\begin{align} \notag
    \text{(P4.2)}: \min_{p_{1},...,p_{N} } \  \  &\sum_{i=1}^{N} \frac{1}{\hat{\sigma}_{i}^2p_{i}} \\
  \text{s.t.}  \quad   &\sum_{i=1}^{N} p_{i} \leq P_0\\
  & p_i \ge 0, \forall i\in \{1,...,N\}.
\end{align}
\end{subequations}
Based on the Karush-Kuhn-Tucker (KKT) conditions, the optimal solution to (P4.2) is presented in the following proposition.
\begin{proposition}\label{proposition_opt_p}
The optimal solution to problem (P4.2) is
\begin{equation}
p_{i}^{\star} = \frac{\hat{\sigma}_{i}^{-1}}{\sum_{i=1}^N\hat{\sigma}_{i}^{-1}}P_0, \quad i \in\{1,...,N\}.
\end{equation}
\begin{IEEEproof}
See Appendix D.
\end{IEEEproof}
\end{proposition}

By combining \eqref{eq:R_X_Q}, \eqref{eq:R_pati}, and Proposition \ref{proposition_opt_p}, we directly have the following proposition.
\begin{proposition}\label{proposition_opt}
The optimal solution to problem (P4) is given by
\begin{equation}\label{eqn:opt:solution:extended}
\mathbf R_x^{\star}= \hat{\mathbf Q} \hat{\mathbf R}_{x}^\text{opt}   \hat{\mathbf Q}^\mathrm{H},
\end{equation}
where \begin{equation}
\hat{\mathbf R}_{x}^{\star}=
\left[
  \begin{array}{cc}
    \frac{\hat{\mathbf \Sigma}_1^{-1} P_0}{\sum_{i=1}^N\hat{\sigma}_{i}^{-1}} &\mathbf 0\\
    \mathbf 0 & \mathbf 0\\
  \end{array}
\right].
\end{equation}
The resultant CRB is
\begin{equation}\label{eq:opt_CRB}
\mathrm{CRB}(\mathbf H)^\text{opt}
=\frac{\sigma_\text{R}^2(\sum_{i=1}^N\hat{\sigma}_{i}^{-1})^2\sum_{i=1}^N\hat{\sigma}_{i}^{-2}}{P_0T}.
\end{equation}
\end{proposition}

In the optimal solution in Proposition \ref{proposition_opt}, SVD is used to decompose the MIMO channel $\bf{G}$ between the AP and the IRS into a set of parallel sensing subchannels, and then a channel amplitude inversion power allocation policy is adopted based on the channel amplitude of each subchannel. Note that the formula in \eqref{eqn:opt:solution:extended} can be viewed as the EVD of the sample coherence matrix $\mathbf R_x^{\star}$. Suppose that $\hat{\mathbf Q}=[\hat{\mathbf q}_1,... ,\hat{\mathbf q}_N]$ in \eqref{eqn:opt:solution:extended}. Then, based on \eqref{eq:coherence_matrix_to_vector}, $\mathbf R_x^{\star}$ in \eqref{eqn:opt:solution:extended} corresponds to that the AP transmits $N$ sensing beams, denoted by $\sqrt{p_i^{\star}}\hat{\mathbf q}_i, i\in\{1, ..., N\}$.

It is interesting to compare the optimal solution $\mathbf R_x^{\star}$ in \eqref{eqn:opt:solution:extended} for CRB minimization with IRS versus the optimal isotropic transmission solution $\mathbf R_x^\text{iso} = P_0\mathbf I_M/M$ without IRS in \cite{9652071}. With $\mathbf R_x^\text{iso}$, the resultant CRB for estimating the target response matrix $\mathbf H$ is given by
\begin{equation}\label{eq:CRB_large}
\begin{split}
\mathrm{CRB}(\mathbf H)^\text{iso}
=&\frac{M\sigma_\text{R}^2(\sum_{i=1}^N\hat{\sigma}_{i}^{-2})^2}{P_0T}\\
\stackrel{(a_{1})}{\ge}&\frac{N\sigma_\text{R}^2(\sum_{i=1}^N\hat{\sigma}_{i}^{-2})^2}{P_0T}\\
\stackrel{(a_{2})}{\ge}& \frac{\sigma_\text{R}^2(\sum_{i=1}^N\hat{\sigma}_{i}^{-1})^2\sum_{i=1}^N\hat{\sigma}_{i}^{-2}}{P_0T}\\
=& \mathrm{CRB}(\mathbf H)^\text{opt},
\end{split}
\end{equation}
where the inequality $(a_{1})$ holds due to the fact that $M\ge N$, and the inequality $(a_{2})$ holds due to the Jensen's inequality. Notice that in \eqref{eq:CRB_large}, $(a_{1})$ and $(a_{2})$ are met with equalities if and only if $M=N$ and $\hat{\sigma}_1 = \hat{\sigma}_2=... = \hat{\sigma}_N$. With $\mathbf R_x^{\star}$ in this case, the radiated signals by the IRS (after reflection) are isotropic, which is thus consistent with the optimality of the isotropic transmission in conventional wireless sensing system without IRS \cite{9652071}.

\section{Numerical Results}
\begin{figure}[t]
\centering
\subfloat[Point target case]{\includegraphics[width=0.23\textwidth]{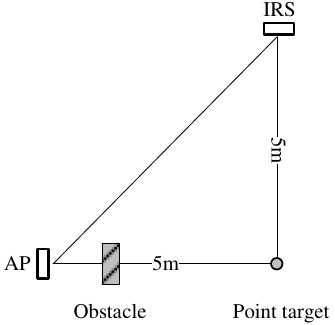}%
}
\subfloat[Extended target case]{\includegraphics[width=0.23\textwidth]{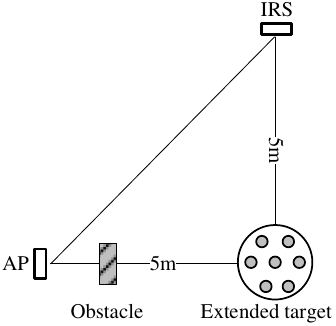}%
}
\caption{Simulation setup.}
\label{Simulation setup}
\end{figure}

This section provides numerical results to evaluate the performance of our proposed joint beamforming design based on CRB minimization. We consider the distance-dependent path loss model, i.e.,
\begin{equation}
L(d)=K_0(\frac{d}{d_0})^{-\alpha_0},
\end{equation}
where $d$ is the distance of the transmission link and $K_0=-30~ \text{dB}$ is the path loss at the reference distance $d_0=1~ \text{m}$, and the path-loss exponent $\alpha_0$ is set as $2.5$ for the AP-IRS and IRS-target links. 
As shown in Fig. \ref{Simulation setup}, the AP and IRS are located at coordinate $(0,0)$ and $(5~\text{m},5~\text{m})$, respectively. For the point target case, the target is located at coordinate $(5~\text{m},0)$ (i.e., the target's DoA is $\theta=0^{\circ}$ w.r.t. the IRS), with a unit RCS. For the extended target case, we model the extended target as $N_s=7$ resolvable scatterers uniformly distributed in a circle with radius $0.5~\text{m}$ and center point $(5~\text{m},5~\text{m})$, with a unit RCS for each scatterer. In this case, the target response matrix is modeled as
\begin{equation}
\mathbf H = \sum_{i=1}^{N_s}\alpha_i\mathbf a(\theta_i)\mathbf a^\mathrm{T} (\theta_i),
\end{equation}
where $\alpha_i$ and $\theta_i$ are the reflection coefficient and the DoA of the $i\text{-th}$ scatterer, respectively. 
We consider the Rician fading channel for the AP-IRS link, i.e.,
\begin{equation}
\mathbf{G}=\sqrt{\frac{\beta_{\mathrm{AI}}}{1+\beta_{\mathrm{AI}}}} \mathbf{G}^{\mathrm{LoS}}+\sqrt{\frac{1}{1+\beta_{\mathrm{AI}}}} \mathbf{G}^{\mathrm{NLoS}},
\end{equation}
where $\beta_{\mathrm{AI}}=0.5$ is the Rician factor, and $\mathbf{G}^{\mathrm{LoS}}$ and $\mathbf{G}^{\mathrm{NLoS}}$ are the LoS and NLoS (Rayleigh fading) components, respectively.  
We set the radar dwell time as $T=256$. We set the spacing between consecutive reflecting elements at the IRS as half wavelength, i.e., $d_\text{IRS}= \lambda_\text{R}/2$ . We also set the noise power at the AP as $\sigma_\text{R}^2 = -120~ \text{dBm}$. 
In the simulation, the results are obtained by averaging over $100$ realizations, unless otherwise mentioned.
\subsection{Point Target Case}
\begin{figure}[t]
    \centering
    \includegraphics[width=0.45\textwidth]{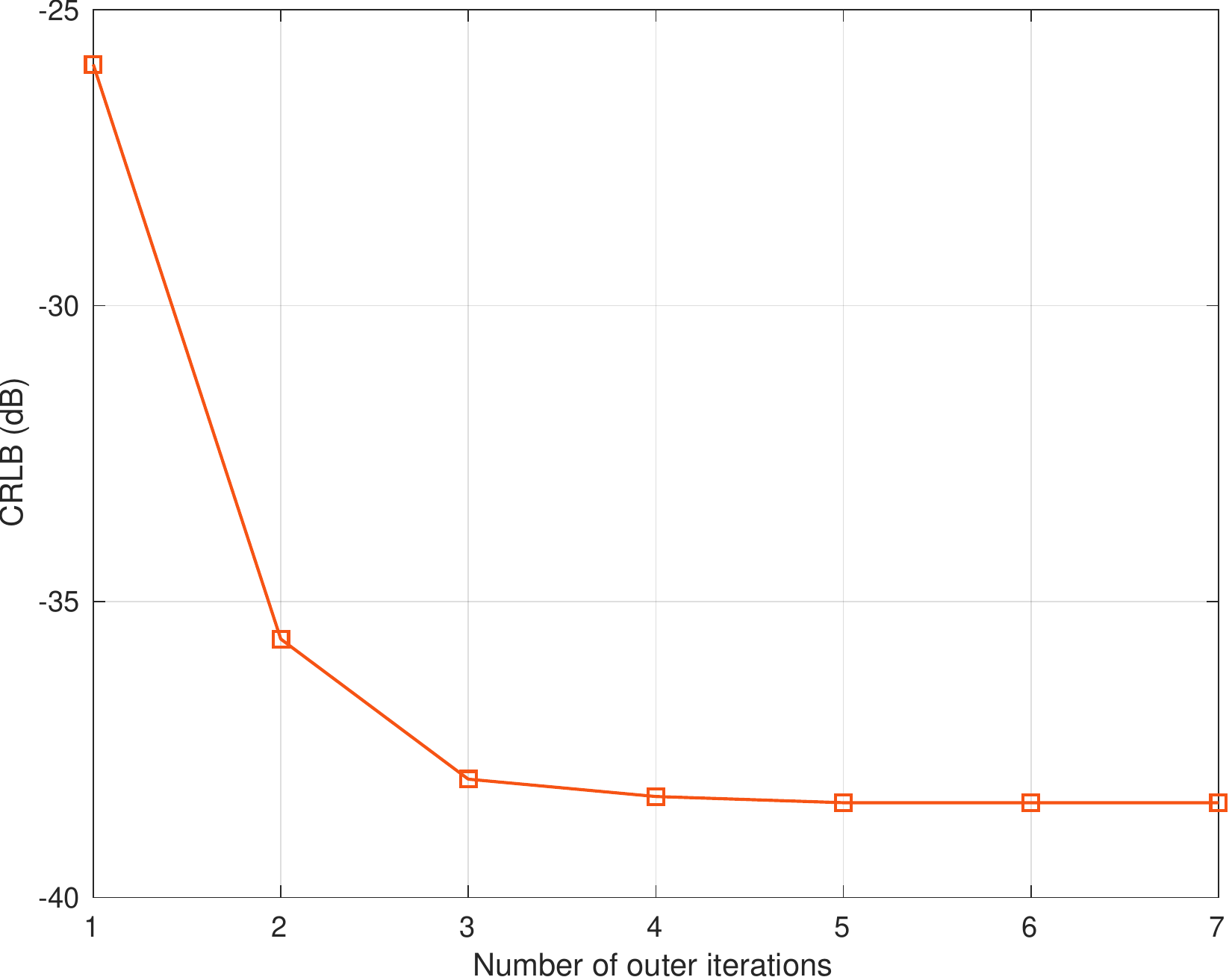}
    \caption{Convergence behavior of the proposed alternating optimization based algorithm for solving problem $\text{(P1)}$, where $P_0 = 30~ \text{dBm}$, $M = 8$, and $N = 8$.}
    \label{convergence}
\end{figure}

First, we consider the point target case. Fig.~\ref{convergence} shows the convergence behavior of our proposed alternating optimization based algorithm for solving problem $\text{(P1)}$, where $P_0 = 30~ \text{dBm}$, $M = 8$, and $N = 8$. It is observed that the proposed alternating optimization based algorithm (Algorithm 1) converges within around $5$ outer iterations, thus validating its effectiveness.

Next, we evaluate the estimation performance of our proposed joint beamforming design based on CRB minimization for IRS-enabled sensing as compared to the following benchmark schemes.

\subsubsection{SNR maximization} We maximize the received SNR of the echo signal at the AP in \eqref{eq:echo_signal}, by jointly optimizing the transmit beamforming at the AP and reflective beamforming at the IRS. First, by applying  maximum-ratio combining at the receiver of the AP, the average SNR of the echo signals is
\begin{equation}\label{eq:SNR}
\begin{split}
    \gamma=&\frac{1}{T}\sum_{t=1}^T\frac{\|\mathbf y(t)\|^2}{\sigma_\text{R}^2}\\
    =&\frac{|\alpha|^2\|\mathbf G^\mathrm{T} \mathbf A \mathbf v\|^2}{\sigma_\text{R}^2}\mathrm{tr}((\mathbf G^\mathrm{T} \mathbf A \mathbf v)^* (\mathbf G^\mathrm{T} \mathbf A \mathbf v)^\mathrm{T} \mathbf R_x).
\end{split}
\end{equation}
Then, with any given reflective beamformer $\mathbf v$, it is well established that maximum-ratio transmission is the optimal transmit beamforming solution, i.e.,
\begin{equation}\label{eq:MRT}
\mathbf R_x^\text{MRT} =\frac{P_0\mathbf G^\mathrm{H} \mathbf A^*  \mathbf v^* \mathbf v^\mathrm{T} \mathbf A^\mathrm{T} \mathbf G}{\|\mathbf G^\mathrm{T} \mathbf A \mathbf v\|^2}.
\end{equation}
By substituting $\mathbf R_x^\text{MRT}$ in \eqref{eq:MRT} into \eqref{eq:SNR}, we have the SNR as
\begin{equation}\label{eq:gamma}
\gamma=\frac{P_0|\alpha|^2}{\sigma_\text{R}^2}\|\mathbf G^\mathrm{T} \mathbf A \mathbf v\|^4.
\end{equation}
Finally, we optimize the reflective beamforming $\mathbf v$ at the IRS to maximize $\|\mathbf G^\mathrm{T} \mathbf A \mathbf v\|^2$ for equivalently maximizing $\gamma$ in \eqref{eq:gamma}, subject to the unit-modulus constraint in \eqref{eq:phase_1}. This is similar to the SNR maximization problem in IRS-enabled multiple-input single-output (MISO) communications that has been studied in \cite{8811733}.

\subsubsection{Reflective beamforming only with isotropic transmission (reflective BF only)} The AP uses the isotropic transmission by transmitting orthonormal signal beams and setting $\mathbf R_x = P_0/M\mathbf I_M$. Then, the reflective beamforming at the IRS is optimized to minimize $\mathrm{CRB}(\theta)$ in \eqref{eq:CRB_1} by solving problem (P3).

\subsubsection{Transmit beamforming only with random phase shifts (transmit BF only)} We consider the random reflecting phase shifts at the IRS, based on which the transmit beamforming at the AP is optimized to minimize $\mathrm{CRB}(\theta)$ in \eqref{eq:CRB} by solving problem (P2).

Furthermore, note that besides the CRB, we also implement the practical MLE method to estimate the target's DoA $\theta$, and accordingly evaluate the MSE as the performance metric for gaining more insights. For brevity, please refer to Appendix E for the details of MLE for the point target case.

\begin{figure}[t]
    \centering
    \includegraphics[width=0.45\textwidth]{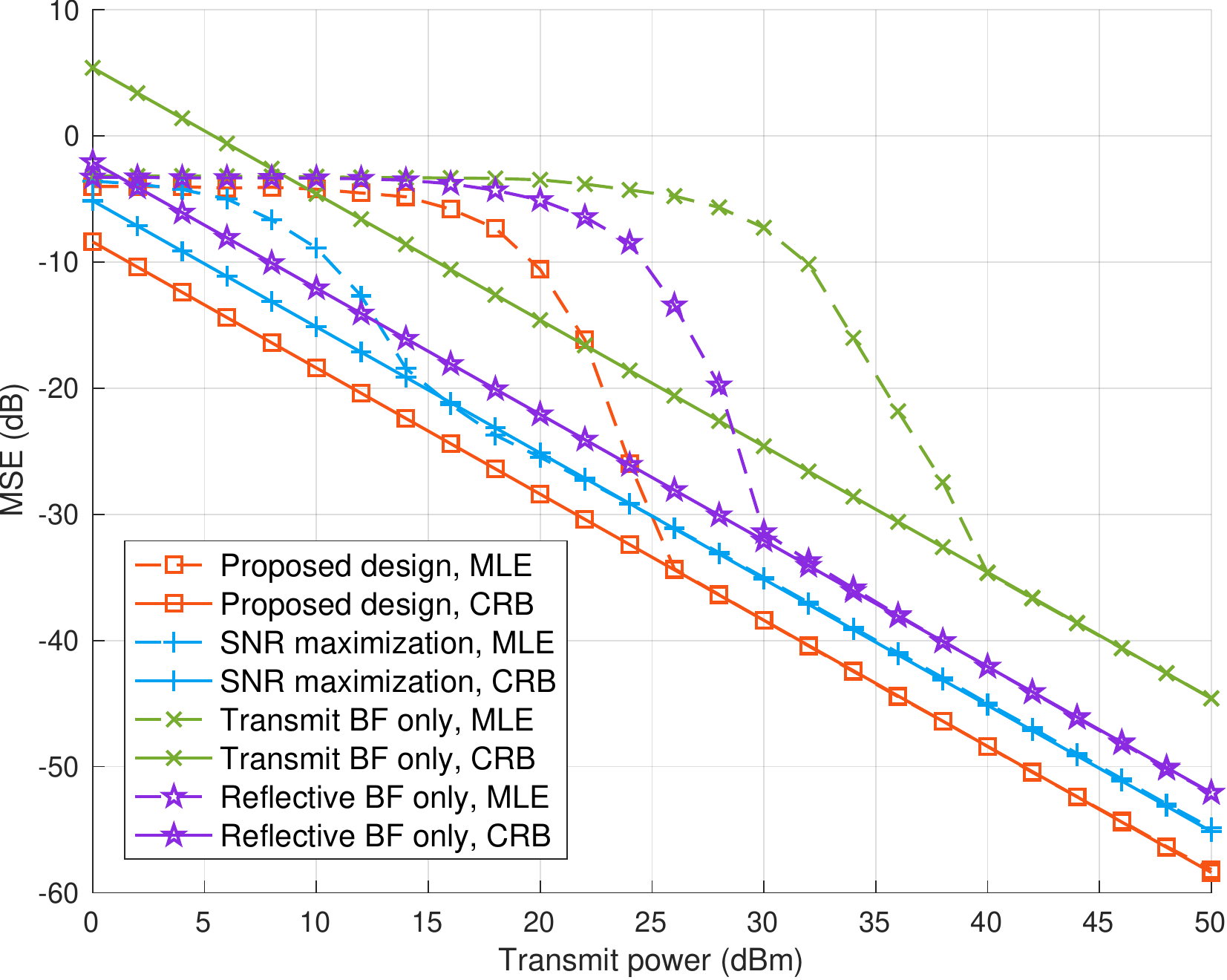}
    \caption{The MSE performance for estimating DoA versus the transmit power $P$ at the AP in point target case, where  $M = 8$ and $N = 8$.}
    \label{correctness}
\end{figure}

Fig.~\ref{correctness} shows the CRB and the MSE with MLE versus the transmit power $P_0$ at the AP. It is observed that the CRB in decibels (dB) is monotonically decreasing in a linear manner w.r.t. the transmit power $P_0$ in dB at the AP. This is because that the $\mathrm{CRB}(\theta)$ in \eqref{eq:CRB} is inversely proportional to the transmit power $P_0$. It is also observed that at the high SNR regime, the derived CRB is identical to the MSE with MLE. This is consistent with the results in \cite{kay1993fundamentals,richards2014fundamentals} and validates the correctness of our CRB derivation. By contrast, at the low SNR regime, the MSE based on the MLE is observed to converge to a constant value when the SNR becomes sufficiently small. This is because that in this case, the estimated DoA with MLE can be any value in the interval of $[-\frac{\pi}{2}, \frac{\pi}{2}]$ with equal probability, which leads to a constant MSE value\cite{richards2014fundamentals}. It is also observed that the proposed CRB minimization scheme achieves the lowest CRB in the whole transmit power regime, which shows the effectiveness of our proposed joint beamforming design in CRB minimization.
Furthermore, when the transmit power is sufficiently high (e.g., $P_0 > 25 ~ \text{dBm}$), the proposed CRB minimization scheme is observed to achieve lower MSE than the three benchmark schemes. When the transmit power is low (e.g., $P_0 < 25 ~ \text{dBm}$), the SNR maximization scheme is observed to achieve lower MSE (with MLE) than the proposed CRB minimization scheme. This is because that in this case, the CRB is not achievable using MLE, and the estimation performance of MLE is particularly sensitive to the power of echo signals, thus making the SNR maximization scheme desirable.

\begin{figure}[t]
    \centering
    \includegraphics[width=0.45\textwidth]{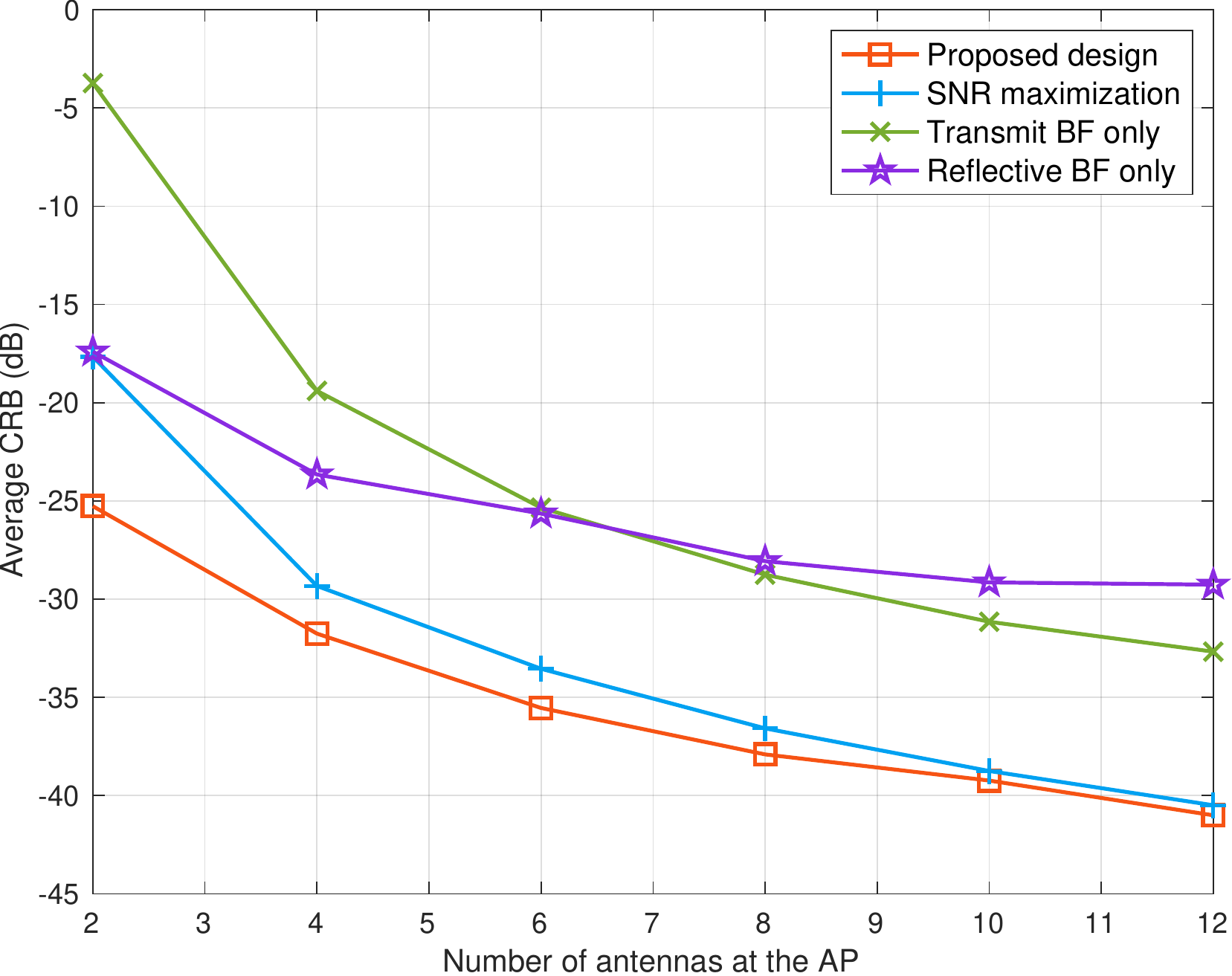}
    \caption{The CRB performance for estimating DoA versus the number of antennas $M$ at the AP in the point target case, where $P_0=30$ dBm and $N=8$.}
    \label{CRLB_point}
\end{figure}

Fig.~\ref{CRLB_point} shows the CRB performance for estimating the DoA versus the number of antennas $M$ at the AP. It is observed that the proposed CRB minimization scheme achieves the lowest CRB in the whole number of antennas regime. It is also observed that with the increasing of the number of antennas $M$ at the AP, the performance gap between the proposed design and reflective beamforming only scheme increases. This shows that the joint beamforming optimizing is particularly important when the number of antennas at the AP becomes large. Furthermore, as $M$ becomes large, the performance gap between the proposed design and SNR maximization scheme is observed to decrease. This shows that the two designs become consistent in this case.  

\subsection{Extended Target Case}
Next, we consider the extended target case. We evaluate the estimation performance of our proposed beamforming design based on CRB minimization as compared to the isotropic transmission with $\mathbf R_x^\text{iso} = P_0\mathbf I_M/M$.

Besides the CRB, we also implement the practical MLE method to estimate the target response matrix $\mathbf H$, and accordingly evaluate the MSE as the performance metric for gaining more insights. For brevity, please refer to Appendix F for the details of MLE for the extended target case. Based on \eqref{eq:vec_data_extended}, the target response matrix estimation problem is a linear estimation model with Gaussian noise. The MLE of target response matrix $\mathbf H$ is a minimum variance unbiased (MVU) estimator of $\mathbf H$ and the MSE for estimating $\mathbf H$ with MLE equals to its CRB \cite{kay1993fundamentals,9747255}.

\begin{figure}[t]
    \centering
    \includegraphics[width=0.45\textwidth]{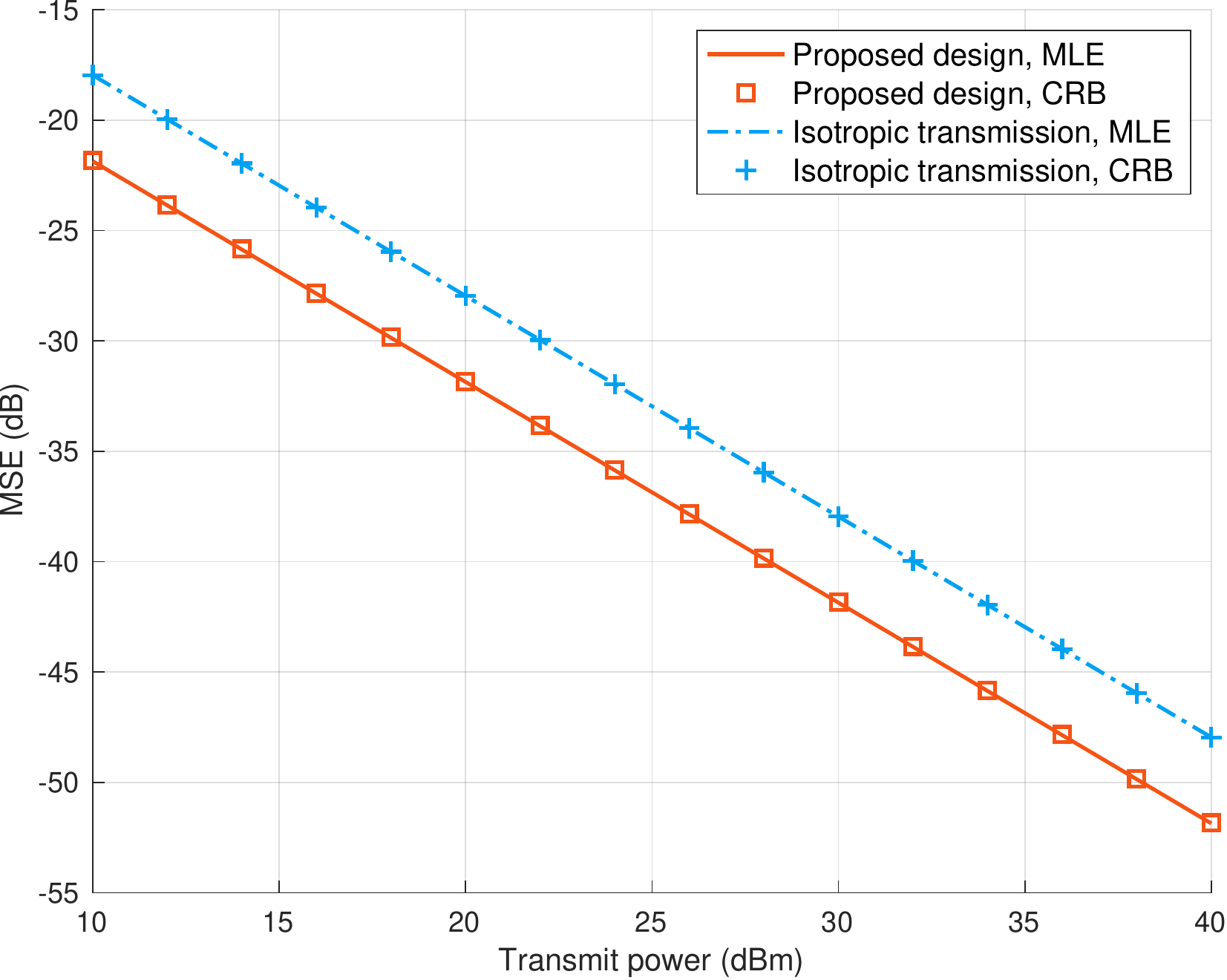}
    \caption{The MSE performance for estimating target response matrix versus the transmit power budget $P$ at the AP in the extended target case, where $M = 8$ and $N = 8$.}
    \label{MSE_extended}
\end{figure}

Fig.~\ref{MSE_extended} shows the CRB and the MSE with MLE versus the transmit power $P_0$ at the AP, where $M = 8$ and $N = 8$. It is observed that similarly as for the point target case, the CRB in dB is monotonically decreasing in a linear manner w.r.t. the transmit power $P_0$ in dB at the AP. This is because that the $\mathrm{CRB}(\mathbf H)$ in \eqref{eq:CRB_extended} is also inversely proportional to the transmit power $P_0$. It is also observed that the derived CRB is identical to the MSE with MLE, which is consistent with the results in \cite{kay1993fundamentals,9747255} and validates the correctness of the CRB derivation. It is also observed that the proposed CRB minimization scheme achieves the lowest CRB in the whole transmit power regime, which shows the effectiveness of our proposed transmit beamforming design for CRB minimization.
\begin{figure}[t]
    \centering
    \includegraphics[width=0.45\textwidth]{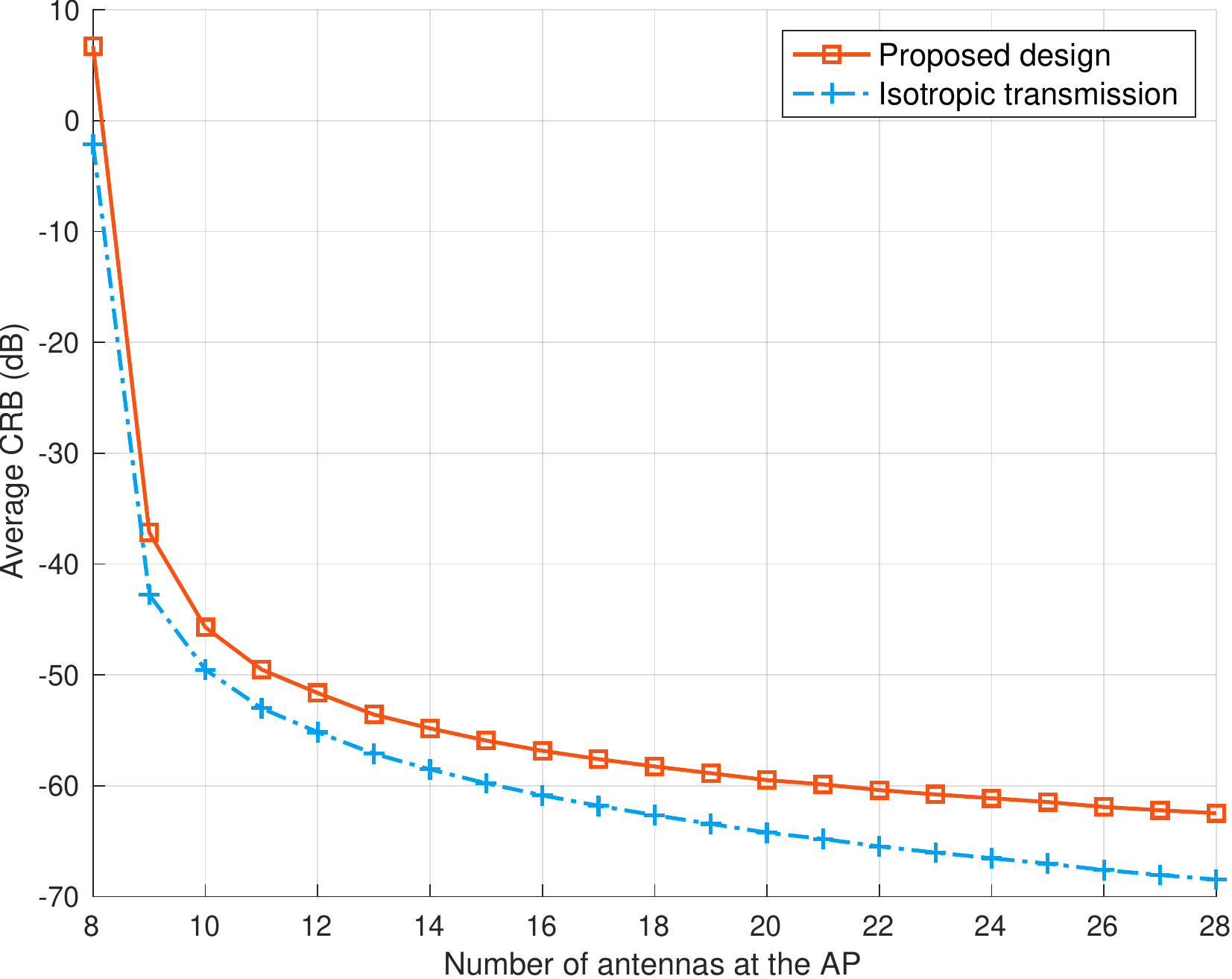}
    \caption{The CRB performance for estimating target response matrix versus the number of antennas $M$ at the AP in the extended target case, where $P_0=30$ dBm and $N=8$.}
    \label{CRLB_extended}
\end{figure}

Fig.~\ref{CRLB_extended} shows the CRB performance for estimating the target response matrix versus the number of antennas $M$ at the AP. It is observed that the proposed CRB minimization design outperforms the benchmarking isotropic transmission scheme in the whole regime of $M$, and the performance gap becomes more significant as $M$ increases.

\section{Conclusion}
This paper considered an IRS-enabled NLoS wireless sensing system consisting of an AP with multiple antennas, an ULA-IRS with multiple reflecting elements, and a target at the NLoS region of the AP. We consider two types of target models, namely the point and extended target models. The AP aimed to estimate the target's DoA and the target response matrix in the two cases, respectively. We first derived the closed-form CRB expressions for parameters estimation. Then, based on the derived results, we optimized the transmit beamforming at the AP and the reflective beamforming at the IRS to minimize the obtained CRB, subject to the maximum power constraint at the AP. Numerical results showed that the proposed CRB minimization scheme achieved the lowest CRB and  MSE with MLE, as compared to other traditional benchmark schemes. We hope that this work can provide new insights in designing IRS-enabled sensing and IRS-enabled ISAC systems based on CRB minimization. How to extend the results to other scenarios (e.g., near-filed IRS-enabled NLoS sensing, distributed target sensing using  multiple widely separated IRSs, and IRS-enabled IASC) is interesting directions worth pursuing in future work.
\appendices

\section{The Derivation of the FIM in \eqref{eq:FIM_partitioned}}
The covariance matrix $\mathbf R_n$ of noise $\tilde{\mathbf n}$ is independent of $\bm \xi$. Therefore, we have $\frac{\partial \mathbf R_n}{\partial \bm \xi_i} = 0, i=1,2,3$, in \eqref{eq:FIM}. Furthermore, 
\begin{equation}
\frac{\partial \tilde{\mathbf u}}{\partial\theta}=\alpha \mathrm{vec}(\dot{\mathbf B}\mathbf X),
\end{equation}
\begin{equation}
\frac{\partial \tilde{\mathbf u}}{\partial\tilde{\bm\alpha}}=[1,j]\otimes  \mathrm{vec}(\mathbf B\mathbf X).
\end{equation}
Accordingly, it follows that
\begin{equation}\label{eq:F_1}
\begin{split}
\tilde{\mathbf F}_{\theta \theta}=&\frac{2}{\sigma_\text{R}^2}\mathrm{Re}\{(\alpha \mathrm{vec}(\dot{\mathbf B}\mathbf X))^\mathrm{H}\alpha \mathrm{vec}(\dot{\mathbf B}\mathbf X)\}\\
=&\frac{2|\alpha|^2}{\sigma_\text{R}^2}\mathrm{Re}\{\mathrm{tr}(\dot{\mathbf B}\mathbf X)^\mathrm{H} (\dot{\mathbf B}\mathbf X)\}\\
=&\frac{2T|\alpha|^2}{\sigma_\text{R}^2}\text{tr}(\dot {\mathbf B}  \mathbf R_x \dot {\mathbf B}^\mathrm{H} ),
\end{split}
\end{equation}
\begin{equation}\label{eq:F_2}
\begin{split}
\tilde{\mathbf F}_{\theta \tilde{\bm\alpha}}=&\frac{2}{\sigma_\text{R}^2}\mathrm{Re}\{(\alpha \mathrm{vec}(\dot{\mathbf B}\mathbf X))^\mathrm{H}[1,j]\otimes  \mathrm{vec}(\mathbf B\mathbf X)\}\\
=&\frac{2}{\sigma_\text{R}^2}\mathrm{Re}\{\alpha^* [1,j]\otimes (\mathrm{vec}(\dot{\mathbf B}\mathbf X)^\mathrm{H} \mathrm{vec}(\mathbf B\mathbf X))\}\\
=&\frac{2}{\sigma_\text{R}^2}\mathrm{Re}\{\alpha^* [1,j] (\mathrm{tr}(\dot{\mathbf B}\mathbf X)^\mathrm{H} \mathbf B\mathbf X)\}\\
=& \frac{2T}{\sigma_\text{R}^2}\mathrm{Re}\{\alpha^*\text{tr}( {\mathbf B}\mathbf R_x \dot {\mathbf B}^\mathrm{H} )[1,j]\},
\end{split}
\end{equation}
\begin{equation}\label{eq:F_3}
\begin{split}
\tilde{\mathbf F}_{\tilde{\bm\alpha} \tilde{\bm\alpha}}=&\frac{2}{\sigma_\text{R}^2}\mathrm{Re}\{([1,j]\!\otimes  \!\mathrm{vec}(\mathbf B\mathbf X))^\mathrm{H}\!([1,j]\!\otimes \!\mathrm{vec}(\mathbf B\mathbf X))\}\\
=&\frac{2}{\sigma_\text{R}^2}\mathrm{Re}\{([1,j]^\mathrm{H} [1,j]) \otimes(\mathrm{vec}(\mathbf B\mathbf X)^\mathrm{H} \mathrm{vec}(\mathbf B\mathbf X))\}\\
=&\frac{2}{\sigma_\text{R}^2}\mathrm{Re}\{([1,j]^\mathrm{H} [1,j]) \mathrm{tr}((\mathbf B\mathbf X)^\mathrm{H} \mathbf B\mathbf X)\}\\
=&\frac{2T}{\sigma_\text{R}^2}\text{tr}( {\mathbf B} \mathbf R_x  {\mathbf B^\mathrm{H}} )\mathbf I_2,
\end{split}
\end{equation}
and the FIM $\tilde{\mathbf F}$ in \eqref{eq:FIM_partitioned} is obtained.

\section{Proof of Proposition $1$}
We obtain the determinant of the FIM $\tilde{\mathbf F}$ in \eqref{eq:FIM_partitioned} to present whether $\tilde{\mathbf F}$ is invertible or not. Based on \eqref{eq:coherence_matrix_to_vector}, \eqref{eq:FIM_partitioned}, \eqref{eq:FIM_1}, \eqref{eq:FIM_2}, and \eqref{eq:FIM_3}, we have
\begin{equation}\label{eq:determinant}
\begin{split}
&\mathrm{det}(\tilde{\mathbf F})\\
=&\frac{8T^3|\alpha|^2}{\sigma^6}\left(\|(\mathbf W \mathbf \Lambda^{1/2})^\mathrm{T} \mathbf b\|^4(\|\dot{\mathbf b} \|^2\|\mathbf b\|^2-|\dot{\mathbf b}^\mathrm{H}  \mathbf b|^2)\right. \\
&+\|\mathbf b\|^4 (\|(\mathbf W \mathbf \Lambda^{1/2})^\mathrm{T} \dot{\mathbf b}\|^2\|(\mathbf W \mathbf \Lambda^{1/2})^\mathrm{T} {\mathbf b}\|^2\\
& \left.-|\dot{\mathbf b}^\mathrm{H}(\mathbf W \mathbf \Lambda^{1/2})^*(\mathbf W \mathbf \Lambda^{1/2})^\mathrm{T} \mathbf b|^2)\right).
\end{split}
\end{equation}

When $\mathrm{rank}(\mathbf G)=1$, the truncated SVD of $\mathbf G$ is expressed as $\mathbf G= \sigma_1 \mathbf s_1 \mathbf q_1^\mathrm{T}$, where $\mathbf s_1$ and $\mathbf q_1$ are the left and right dominant singular vectors, and $\sigma_1$ denotes the dominant singular value. It follows that
\begin{equation}
\mathbf b=\sigma_1\mathbf q_1 \mathbf s_1^\mathrm{T}\mathbf{A}\mathbf v,
\end{equation}
\begin{equation}
\dot{\mathbf b} =j 2\pi \frac{d_\text{IRS}}{\lambda_\text{R}}\cos\theta \sigma_1\mathbf q_1 \mathbf s_1^\mathrm{T}\mathbf{A} \mathbf D\mathbf v,
\end{equation}
\begin{equation}
(\mathbf W \mathbf \Lambda^{1/2})^\mathrm{T}\mathbf b=(\mathbf W \mathbf \Lambda^{1/2})^\mathrm{T} \sigma_1\mathbf q_1 \mathbf s_1^\mathrm{T}\mathbf{A}\mathbf v,
\end{equation}
\begin{equation}
(\mathbf W \mathbf \Lambda^{1/2})^\mathrm{T}\dot{\mathbf b}=(\mathbf W \mathbf \Lambda^{1/2})^\mathrm{T} j 2\pi \frac{d_\text{IRS}}{\lambda_\text{R}}\cos\theta \sigma_1\mathbf q_1 \mathbf s_1^\mathrm{T}\mathbf{A} \mathbf D\mathbf v.
\end{equation}
It is clear that  $\dot{\mathbf b}$ and $\mathbf b$ are aligned with each other, i.e.,
\begin{equation}
\dot{\mathbf b}=\frac{j 2\pi d_\text{IRS}\cos\theta \mathbf s_1^\mathrm{T}\mathbf{A} \mathbf D\mathbf v}{\lambda_\text{R}\mathbf s_1^\mathrm{T}\mathbf{A}\mathbf v}\mathbf b,
\end{equation}
such that
\begin{equation}\label{eq:align_1}
|\dot{\mathbf b}^\mathrm{H}  \mathbf b|^2=\|\dot{\mathbf b} \|^2\|\mathbf b\|^2.
\end{equation}
Similarly, as $(\mathbf W \mathbf \Lambda^{1/2})^\mathrm{T}\dot{\mathbf b}$ and $(\mathbf W \mathbf \Lambda^{1/2})^\mathrm{T}\mathbf b$ are aligned with each other, we have
\begin{equation}\label{eq:align_2}
\begin{split}
&|\dot{\mathbf b}^\mathrm{H}(\mathbf W \mathbf \Lambda^{1/2})^*(\mathbf W \mathbf \Lambda^{1/2})^\mathrm{T} \mathbf b|^2\\
=&\|(\mathbf W \mathbf \Lambda^{1/2})^\mathrm{T} \dot{\mathbf b}\|^2\|(\mathbf W \mathbf \Lambda^{1/2})^\mathrm{T} {\mathbf b}\|^2.
\end{split}
\end{equation}
Based on \eqref{eq:determinant}, \eqref{eq:align_1}, and \eqref{eq:align_2},  $\mathrm{det}(\tilde{\mathbf F})=0$, which implies that the FIM $\tilde{\mathbf F}$ is not invertible. Accordingly, the CRB in \eqref{eq:CRB_1} is unbounded.

Next, suppose that $\mathrm{rank}(\mathbf G)>1$, in this case  $\dot{\mathbf b}$ and $\mathbf b$ are not aligned with each other. According to the Cauchy-Schwarz inequality, we have
\begin{equation}\label{eq:align_3}
|\dot{\mathbf b}^\mathrm{H}  \mathbf b|^2<\|\dot{\mathbf b} \|^2\|\mathbf b\|^2.
\end{equation}
Similarly, as $(\mathbf W \mathbf \Lambda^{1/2})^\mathrm{T}\dot{\mathbf b}$ and $(\mathbf W \mathbf \Lambda^{1/2})^\mathrm{T}\mathbf b$ are not aligned with each other, we have
\begin{equation}\label{eq:align_4}
\begin{split}
&|\dot{\mathbf b}^\mathrm{H}(\mathbf W \mathbf \Lambda^{1/2})^*(\mathbf W \mathbf \Lambda^{1/2})^\mathrm{T} \mathbf b|^2\\
<&\|(\mathbf W \mathbf \Lambda^{1/2})^\mathrm{T} \dot{\mathbf b}\|^2\|(\mathbf W \mathbf \Lambda^{1/2})^\mathrm{T} {\mathbf b}\|^2.
\end{split}
\end{equation}
Based on \eqref{eq:determinant}, \eqref{eq:align_3}, and \eqref{eq:align_4}, $\mathrm{det}(\tilde{\mathbf F})\neq0$, which implies that the FIM $\tilde{\mathbf F}$ is invertible. Accordingly, the CRB in \eqref{eq:CRB_1} is bounded. 

\section{The Derivation of the FIM in \eqref{eq:FIM_partitioned_extended} }
As the covariance matrix $\mathbf R_n$ of noise $\hat{\mathbf n}$ is independent with $\bm \zeta$, we have $\frac{\partial \mathbf R_n}{\partial {\bm \zeta_i}} = 0, i=1,...,2N^2$, in \eqref{eq:FIM}. Furthermore, we have
\begin{equation}
\frac{\partial \hat{\mathbf u}}{\partial {\mathbf h}_\text{R}}=\mathbf X^\mathrm{T}\mathbf{G}^\mathrm{T}\mathbf{\Phi}^\mathrm{T}\otimes \mathbf{G}^\mathrm{T}\mathbf{\Phi}^\mathrm{T},
\end{equation}
\begin{equation}
\frac{\partial \hat{\mathbf u}}{\partial\mathbf h_\text{I}}=j \mathbf X^\mathrm{T}\mathbf{G}^\mathrm{T}\mathbf{\Phi}^\mathrm{T}\otimes \mathbf{G}^\mathrm{T}\mathbf{\Phi}^\mathrm{T}.
\end{equation}
Accordingly, it follows that
\begin{equation}
\begin{split}
\hat{\mathbf F}_{\mathbf h_\text{R} \mathbf h_\text{R}}=&\frac{2T}{\sigma_\text{R}^2}\mathrm{Re}\{(\mathbf X^\mathrm{T}\mathbf{G}^\mathrm{T}\mathbf{\Phi}^\mathrm{T}\otimes \mathbf{G}^\mathrm{T}\mathbf{\Phi}^\mathrm{T})^\mathrm{H} \cdot\\
& (\mathbf X^\mathrm{T}\mathbf{G}^\mathrm{T}\mathbf{\Phi}^\mathrm{T}\otimes \mathbf{G}^\mathrm{T}\mathbf{\Phi}^\mathrm{T})\}\\
=&\frac{2T}{\sigma_\text{R}^2}\mathrm{Re}\{(\mathbf{\Phi}^*\mathbf{G}^*\mathbf R_X^\mathrm{T}\mathbf{G}^\mathrm{T}\mathbf{\Phi}^\mathrm{T})\otimes (\mathbf{\Phi}^*\mathbf{G}^*\mathbf{G}^\mathrm{T}\mathbf{\Phi}^\mathrm{T})\}.
\end{split}
\end{equation}
Similarly, we have
\begin{equation}
\hat{\mathbf F}_{\mathbf h_\text{I} \mathbf h_\text{I}}=\frac{2T}{\sigma_\text{R}^2}\mathrm{Re}\{(\mathbf{\Phi}^*\mathbf{G}^*\mathbf R_X^\mathrm{T}\mathbf{G}^\mathrm{T}\mathbf{\Phi}^\mathrm{T})\otimes (\mathbf{\Phi}^*\mathbf{G}^*\mathbf{G}^\mathrm{T}\mathbf{\Phi}^\mathrm{T})\},
\end{equation}
\begin{equation}
\begin{split}
&\hat{\mathbf F}_{\mathbf h_\text{I} \mathbf h_\text{R}}=-\hat{\mathbf F}_{\mathbf h_\text{R} \mathbf h_\text{I}}\\
=&\frac{2T}{\sigma_\text{R}^2}\mathrm{Im}\{(\mathbf{\Phi}^*\mathbf{G}^*\mathbf R_X^\mathrm{T}\mathbf{G}^\mathrm{T}\mathbf{\Phi}^\mathrm{T}) \otimes (\mathbf{\Phi}^*\mathbf{G}^*\mathbf{G}^\mathrm{T}\mathbf{\Phi}^\mathrm{T})\}.
\end{split}
\end{equation}

\section{Proof of Proposition $3$}
The Lagrangian of problem (P4.2) is
\begin{equation}
\mathcal{L} = \sum_{i=1}^{N} \frac{1}{\hat{\sigma}_{i}^2p_{i}} + \mu (\sum_{i=1}^{N} p_{i} - P_0) -  \sum_{i=1}^{N}\eta_i p_{i},
\end{equation}
where $\mu$ and $\eta_i, i\in\{ 1,...,N\}$, denote the dual variables. Accordingly, the KKT conditions are given by
\begin{equation}
\frac{\partial\mathcal{L}}{\partial p_{i}}= -\frac{1}{\hat{\sigma}_{i}^2p_{i}^2}+\mu-\eta_i=0, \forall i \in\{1,..,N\},
\end{equation}
\begin{equation}
\mu (\sum_{i=1}^{N} p_{i} - P_0)= 0,\mu\ge 0,\sum_{i=1}^{N} p_{i} \le P_0,
\end{equation}
\begin{equation}
\eta_i p_{i}= 0, \eta_i\ge 0, p_{i} > 0, \forall i \in\{1,..,N\}.
\end{equation}
Then the closed-form solution to problem (P4.2) is
\begin{equation}
p_{i}^{\star}=\frac{\hat{\sigma}_{i}^{-1}}{\sum_{i=1}^N\hat{\sigma}_{i}^{-1}}P_0, \quad i \in\{1,...,N\}.
\end{equation}

\section{MLE for Estimating DoA with Point Target}
Based on \eqref{eq:vec_data}, the vectorized received signal at the AP is rewritten as
\begin{equation}
\tilde{\mathbf y}=\alpha \mathbf d (\theta) + \tilde{\mathbf n},
\end{equation}
where $\mathbf d(\theta) =\mathrm{vec}(\mathbf B \mathbf X)$.
The likelihood function of $\tilde{\mathbf y}$ given $\bm \xi$ is
\begin{equation}
 f_{\tilde{\mathbf y}}(\tilde{\mathbf y}; \bm \xi)=\frac{1}{(\pi\sigma_\text{R}^2)^{MT}} \exp (-\frac{1}{\sigma_\text{R}^2}\|\tilde{\mathbf y}-\alpha \mathbf d(\theta)\|^2).
\end{equation}
In this case, maximizing  $ f_{\tilde{\mathbf y}}(\tilde{\mathbf y}; \bm \xi)$ is equivalent to minimizing $\|\tilde{\mathbf y}-\alpha \mathbf d(\theta)\|^2$.
Hence, the MLE of $\theta$ and $\alpha$ is given by
\begin{equation}\label{eq:ML}
(\theta_\text{MLE},{\mathbf \alpha}_\text{MLE})=\arg \min_{\theta,\mathbf \alpha} \|\tilde{\mathbf y}-\alpha \mathbf d (\theta)\|^2.
\end{equation}
Note that under any given $\theta$, the MLE of $\alpha$ is obtained as
\begin{equation}\label{eq:Ml_alpha}
\alpha_\text{MLE}=(\mathbf d^\mathrm{H}(\theta) \mathbf d(\theta))^{-1}\mathbf d^\mathrm{H}(\theta)\tilde{\mathbf y}=\frac{\mathbf d^\mathrm{H}(\theta)\tilde{\mathbf y}}{\|\mathbf d(\theta)\|^2}.
\end{equation}
By substituting \eqref{eq:Ml_alpha} back into \eqref{eq:ML}, we have
\begin{equation}\label{eq:L}
\begin{split}
\|\tilde{\mathbf y}-\alpha_\text{MLE} \mathbf d (\theta)\|^2=&\|\tilde{\mathbf y}\|^2-\frac{|\mathbf d^\mathrm{H}(\theta)\tilde{\mathbf y}|^2}{\|\mathbf d(\theta)\|^2}\\
=&\|\mathrm{vec}(\mathbf Y)\|^2-\frac{|\mathrm{vec}(\mathbf B\mathbf X)^\mathrm{H}\mathrm{vec}(\mathbf Y)|^2}{\mathrm{vec}(\mathbf B\mathbf X)^\mathrm{H}\mathrm{vec}(\mathbf B\mathbf X)}\\
=&\|\mathrm{vec}(\mathbf Y)\|^2-\frac{|\mathrm{tr}(\mathbf B^\mathrm{H}\mathbf Y\mathbf X^\mathrm{H})|^2}{\mathrm{tr}((\mathbf B\mathbf X)^\mathrm{H}\mathbf B\mathbf X)}\\
=&\|\mathrm{vec}(\mathbf Y)\|^2-\frac{|\mathbf b^\mathrm{H}\mathbf Y\mathbf X^\mathrm{H}\mathbf b^*|^2}{T\mathrm{tr}(\mathbf B\mathbf R_x\mathbf B^\mathrm{H})}\\
=&\|\mathrm{vec}(\mathbf Y)\|^2-\frac{|\mathbf b^\mathrm{H}\mathbf Y\mathbf X^\mathrm{H}\mathbf b^*|^2}{T\|\mathbf b\|^2\mathbf b^\mathrm{H}\mathbf R_x^\mathrm{T}\mathbf b}.
\end{split}
\end{equation}
As a result, by substituting \eqref{eq:L} into \eqref{eq:ML}, the MLE of $\theta$ becomes
\begin{equation}
\theta_\mathrm{MLE}= \arg \max_\theta \frac{|\mathbf b^\mathrm{H}\mathbf Y\mathbf X^\mathrm{H}\mathbf b^*|^2}{T\|\mathbf b\|^2\mathbf b^\mathrm{H}\mathbf R_x^\mathrm{T}\mathbf b},
\end{equation}
which can be obtained via exhaustive search over $[-\frac{\pi}{2}, \frac{\pi}{2}]$.

\section{MLE for Estimating Target Response Matrix with Extended Target}
Based on \eqref{eq:vec_data_extended}, the vectorized received signal at the AP is rewritten as
\begin{equation}
\hat{\mathbf y}=\mathbf E \mathbf h + \hat{\mathbf n},
\end{equation}
where $\mathbf E = \mathbf X^\mathrm{T}\mathbf{G}^\mathrm{T}\mathbf{\Phi}^\mathrm{T}\otimes \mathbf{G}^\mathrm{T}\mathbf{\Phi}^\mathrm{T}$. The likelihood function of $\tilde{\mathbf y}$ given $\mathbf h$ is
\begin{equation}
 f_{\hat{\mathbf y}}(\hat{\mathbf y}; \mathbf h)=\frac{1}{(\pi\sigma_\text{R}^2)^{MT}} \exp (-\frac{1}{\sigma_\text{R}^2}\|\hat{\mathbf y}-\mathbf E \mathbf h\|^2).
\end{equation}
In this case, maximizing  $  f_{\hat{\mathbf y}}(\hat{\mathbf y}; \mathbf h)$ is equivalent to minimizing $\|\hat{\mathbf y}-\mathbf E \mathbf h\|^2$.
Hence, the MLE of $\mathbf h$ is given by
\begin{equation}\label{eq:ML_h}
{\mathbf h}_\text{MLE}=\arg \min_{\mathbf h} \|\hat{\mathbf y}-\mathbf E \mathbf h\|^2 = (\mathbf E^\mathrm{H}\mathbf E)^{-1}\mathbf E^\mathrm{H} \tilde{\mathbf y}.
\end{equation}

\ifCLASSOPTIONcaptionsoff
  \newpage
\fi

\bibliographystyle{IEEEtran}
\bibliography{IEEEabrv,mybibfile}

\end{document}